\documentclass[aps,prl,floatfix,twocolumn,groupedaddress]{revtex4}
\usepackage{epstopdf}
\usepackage{times}
\usepackage{amssymb,amsmath}
\usepackage{array}
\usepackage{multirow}
\usepackage{color}
\usepackage{setspace}
\usepackage{changes}
	\definechangesauthor[name={Per cusse}, color=orange]{per}

\usepackage{hyperref}
\hypersetup{hypertex=true,
    colorlinks=true,
    linkcolor=blue,
    anchorcolor=blue,
    citecolor=blue}
    
\usepackage{braket}
\usepackage{booktabs}
\usepackage{color}
\usepackage{threeparttable}
\usepackage{multirow}
\usepackage{mathtools}

\makeatletter

\begin{document}

\title[]{A measurement-device-independent quantum key distribution network using optical frequency comb}

\author{Wenhan Yan$^{1,\dagger}$, Xiaodong Zheng$^{1,\dagger}$, Wenjun Wen$^{1}$, Liangliang Lu$^{2,4}$, Yifeng Du$^{1}$, Yanqing Lu$^{1}$, Shining Zhu$^{1}$, Xiao-Song Ma$^{1,3,4,\ast}$}
\affiliation{
$^1$~National Laboratory of Solid-State Microstructures, Collaborative Innovation Center of Advanced Microstructures, School of Physics, Nanjing University, Nanjing 210093, China\\
$^2$~Key Laboratory of Optoelectronic Technology of Jiangsu Province, School of Physical Science and Technology, Nanjing Normal University, Nanjing 210023, China\\
$^3$~Synergetic Innovation Center of Quantum Information and Quantum Physics, University of Science and Technology of China, Hefei, 230026, China\\
$^4$~Hefei National Laboratory, Hefei 230088, China\\ 
$^{\dagger}$These authors contributed equally to this work\\
$^{\ast}$e-mail: xiaosong.ma@nju.edu.cn
}
\date{\today}

\begin{abstract}	
Quantum key distribution (QKD), which promises secure key exchange between two remote parties, is now moving toward the realization of scalable and secure QKD networks (QNs). Fully connected, trusted node-free QNs have been realized based on entanglement distribution, in which the low key rate as well as the large overhead makes their practical deployment and application challenging. Here, we propose and experimentally demonstrate a fully connected multi-user QKD network based on a wavelength-multiplexed measurement-device-independent (MDI) QKD protocol. By combining this novel protocol with integrated optical frequency combs, we achieve an average secure key rate of 267 bits per second for about 30 dB of link attenuation per user pair---more than three orders of magnitude higher than previous entanglement-based works. More importantly, we realize secure key sharing between two different pairs of users simultaneously, which requires four-photon detection and is not possible with the previous two-photon entanglement distribution. Our work paves the way for the realization of large-scale QKD networks with full connectivity and simultaneous communication capability among multiple users.
\end{abstract}

\maketitle
\onecolumngrid
\setstretch{1.667}
\section*{Introduction}
Quantum key distribution (QKD)~\cite{BB84,Gisin2002,Scarani2009,xu2019,Pirandola2020,PhysRevLett.67.661} allows secret key exchange between two remote parties with unconditional information security guaranteed by the laws of quantum physics. In practice, QKD implementations still deviate from the ideal description due to the imperfection of realistic devices that can open side channels for eavesdroppers~\cite{Brassard2000,PhysRevA.75.032314,Xu_2010,Tang2013,Makarov2006,PhysRevA.61.052304,zhao2008,gerhardt2011full,Lydersen_2011,Weier_2011}. Measurement-device-independent (MDI) QKD~\cite{braunstein2012side,Lo2012,ma2012alternative,da2013proof,rubenok2013real,liu2013experimental,PhysRevLett.122.160501,PhysRevLett.112.190503} removes all side channels on the detection side. Several efforts have been made to improve its practical performance including long distance~\cite{tang2014,Yin2016}, high key rate~\cite{comandar2016,Woodward2021}, and field tests~\cite{tang2015,PhysRevX.6.011024,yuan2020}. Compared with the recent twin-field (TF) QKD~\cite{lucamarini2018,Chen2021TF,Pittaluga2021,Wang2022,Liu2021tf,li2022twin,zhou2023twin,liu2023e,PhysRevLett.124.070501} which can overcome the rate-distance limit, MDI-QKD requires no global phase tracking of distant sources and less phase stabilization of the optical path. Therefore, it is considered as a practical upgrade for next-generation QKD systems. Recently, chip-based MDI-QKD including high-speed transmitter chip~\cite{Wei2020,Semenenko2020} and on-chip untrusted relay~\cite{cao2020chip,Zheng2021} has attracted much attention due to its scalability of device fabrication, small footprint and low cost.

To date, various topologies of QKD networks (QNs) have been developed in large domains~\cite{Elliott_2002,Peev_2009,chen2009field,sasaki2011field,chen2021,proietti2021,huang2023fully}. The goal of QN is to provide wide area connectivity and high key rate without compromising security. The point-to-multipoint quantum access network based on Bennett-Brassard-1984 (BB84) protocol~\cite{frohlich2013quantum}, allows multiple users to share receivers or sources and has been realized in configurations with passive optical splitters and active optical switches that establish a temporary quantum channel between two particular users at a time. MDI-QKD is the time-reversal of entanglement-based QKD and it is naturally suitable for extension to multi-user star-type metropolitan QNs. In 2016, the first MDI-QN with an optical switch~\cite{PhysRevX.6.011024} was realized, where a specific pair of users can exchange keys at the same time. Recently, a fully connected trusted node-free QN scheme based on wavelength-multiplexed entanglement distribution has been proposed and realized~\cite{wengerowsky2018}. This pioneering work has triggered the recent growing interests in combining entanglement-based Bennett-Brassard-Mermin-1992 (BBM92) QKD with classical wavelength division multiplexing (WDM) technology~\cite{joshi2020,liu202240,Wen2022}. However, the low key rate as well as the large overhead makes the practical deployment and application of this scheme challenging.

In this work, we propose and experimentally demonstrate a fully connected multi-user MDI-QN where all users in the network can exchange secure keys. The novelties of our work are as follows: 1. From the protocol perspective, we go beyond the traditional point-to-point QKD architecture (including BB84, BBM92, and MDI protocols) and realize a fully connected QN without trusted nodes. Moreover, our scheme allows the experimental realization of genuine secure key distribution among different users simultaneously, which is not realized in the fully connected entanglement-based QNs~\cite{wengerowsky2018,joshi2020,liu202240,Wen2022}. 2. From the hardware perspective, we use the dissipative Kerr soliton (DKS) optical frequency comb~\cite{herr2014temporal,kippenberg2018d,wang2020quantum} generated from integrated silicon nitride ($Si_3N_4$) microring resonator (MRR) as the light source, which provides multiple highly coherent frequency lines in a wide bandwidth ($\sim$ 120nm), covering the S, C, and L bands with 100~GHz frequency spacing, thus greatly reducing the resource overhead. 3. From the system performance and results perspective, we obtain an average secure key rate of 267 bps with about 30 dB of link attenuation, which is about three orders of magnitude higher than previous fully connected trusted node-free QN demonstrations~\cite{wengerowsky2018,joshi2020,liu202240,Wen2022}. More importantly, we achieve simultaneous secure key exchange between two different pairs of users with a rate of 0.1 Hz under about 30 dB of attenuation per user pair, which is not achievable in the previous fully connected entanglement-based QNs with time-division multiplexing (TDM) method.~\cite{wengerowsky2018,joshi2020,liu202240,Wen2022}.

\section*{Results}

\begin{figure*}[htbp]
\begin{center}
    \includegraphics[width=0.8\textwidth]{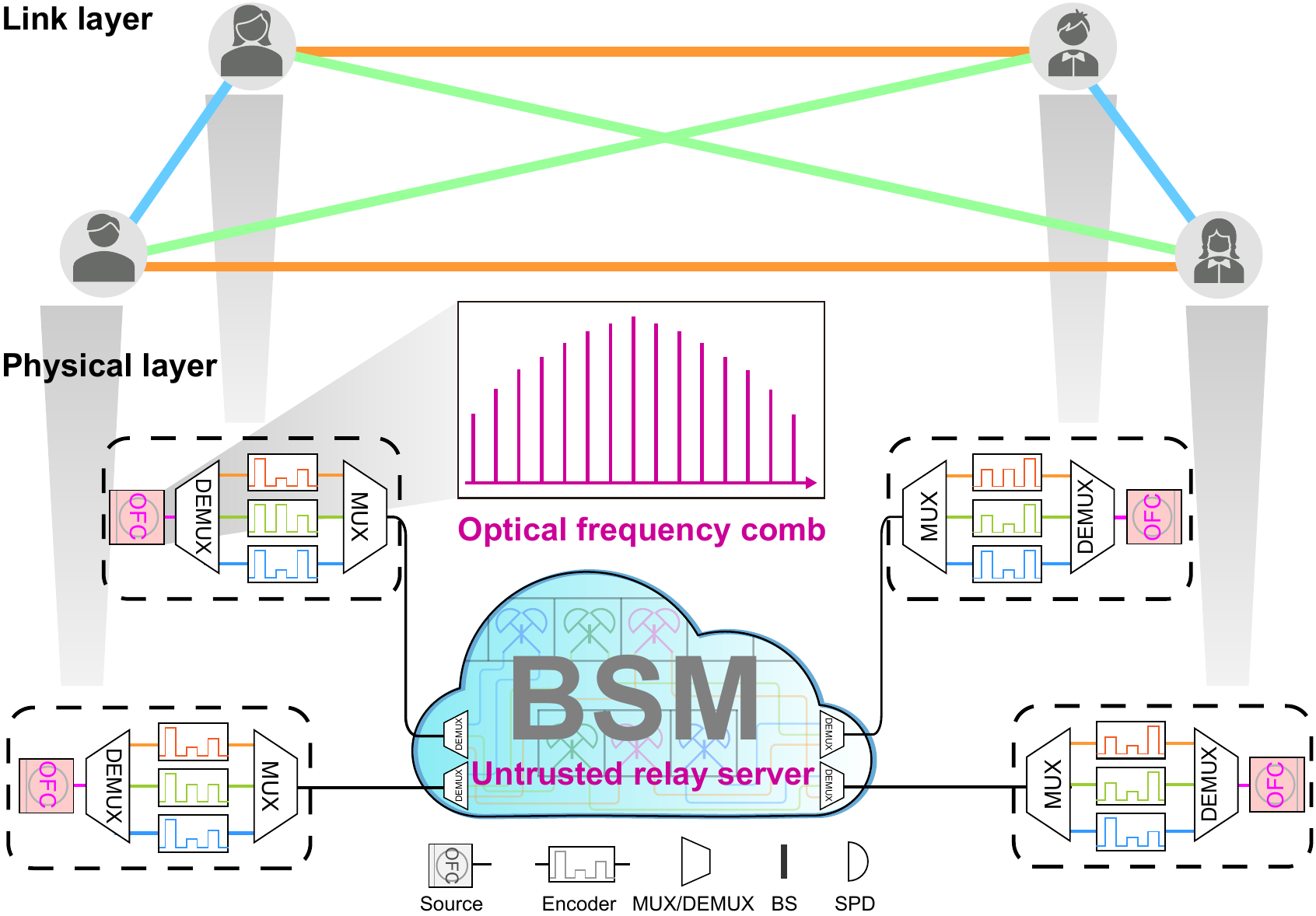}
    \caption{\label{Fig1} Architecture of the fully connected measurement-device-independent quantum key distribution network (MDI-QN). At the link layer, each independent communication link establishes the correlation between two remote users using Bell-state measurement (BSM). The colors represent the wavelengths of the encoded qubits. At the physical layer, each user of the MDI-QN uses a transmitter containing an optical frequency comb (OFC), a wavelength demultiplexer (DEMUX), encoders, and a wavelength multiplexer (MUX) to randomly generate encoded qubits with different wavelengths. At the central untrusted relay server, the encoded qubits are routed by the DEMUXs to different BSM modules to establish the shared secret keys among the users in this MDI-QN.} 
\end{center}
\end{figure*}

The schematic of our four-user fully connected MDI-QN is shown in Fig. \ref{Fig1}. The link layer of the fully connected QN is formed among four users (A, B, C, D) by using six communication links. The colors of different links represent different wavelength channels (top panel of Fig. \ref{Fig1}). The physical layer is shown in the bottom panel of Fig. \ref{Fig1}. Each user employs a transmitter including an optical frequency comb (OFC), a wavelength demultiplexer (DEMUX), encoders, and a wavelength multiplexer (MUX) to randomly generate encoded qubits of different wavelengths. These encoded qubits are then sent over a single optical fiber to the untrusted relay server, which contains multiple DEMUXs and Bell-state measurement (BSM) modules, including beam splitters and single-photon detectors (SPDs). Our scheme can be extended to a $n$-user fully connected MDI-QN. If the number of users $n$ is even (odd), we need $n-1$ ($n$) frequency lines, which can be competently provided by a single DKS optical frequency comb with its broad frequency spectrum. At the central untrusted relay server, $n(n-1)/2$ BSM modules are required. Although cryogenically cooled SPDs are resource intensive, the recently developed waveguide-integrated superconducting SPD array~\cite{Beutel2021,haussler2022scaling,Zheng2021} provides an excellent solution where a single cryostat can accommodate multiple BSM modules. Compared to the entanglement-based QNs~\cite{wengerowsky2018,joshi2020,liu202240,Wen2022}, a significant amount of overhead is saved on the detection side, since there each user needs SPDs along with the cryostat that are spatially separated.

\begin{figure*}[htbp]
\begin{center}
    \includegraphics[width=1\textwidth]{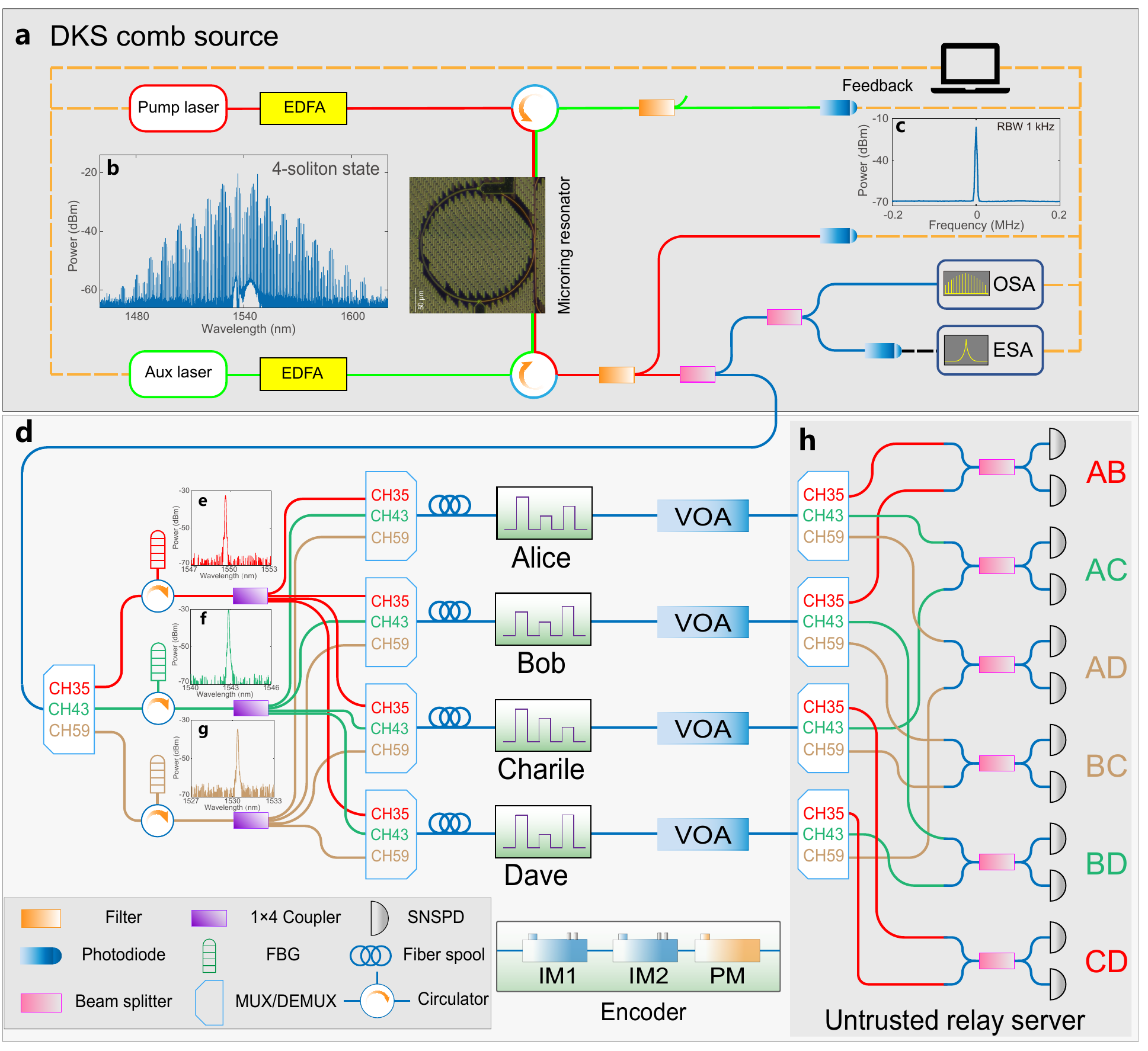}
    \caption{\label{Fig2} Experimental setup of the four-user fully connected MDI-QN. \textbf{a}, An integrated silicon nitride ($Si_3N_4$) microring resonator generates the DKS comb via a dual-driven scheme involving two pump lasers and a feedback loop. The insets (\textbf{b}, \textbf{c}) are the optical spectrum and the electrical beat note of the 4-soliton state acquired by an optical spectrum analyzer (OSA) and an electrical spectrum analyzer (ESA), respectively. \textbf{d}, Three comb lines separated by WDMs act as light sources for four users. The insets (\textbf{e}, \textbf{f}, \textbf{g}) are the optical spectra of selected comb lines measured by the OSA. A combination of three 1x4 couplers and four MUXs is used to route the light of different wavelengths to the users. Each user possesses an encoder consisting of two intensity modulators (IMs) and a phase modulator (PM) that generates time-bin encoded qubits. The following variable optical attenuators (VOAs) are used to attenuate the pulses to the single-photon level and simulate channel loss.\textbf{h}, Encoded qubits with different wavelengths are routed to different BSM modules by DEMUXs at the untrusted relay server. All photons are detected by superconducting nanowire single-photon detectors (SNSPDs). See text for details.}
\end{center}
\end{figure*}

We perform a proof-of-principle demonstration of a four-user fully connected MDI-QN. The layout of our experimental setup is shown in Fig. \ref{Fig2}. We use an integrated $Si_3N_4$ microring resonator (MRR) with a waveguide cross section of about 1.6~$\mu$m$\times$0.8~$\mu$m for DKS optical comb generation (Ligentec). The radius of the MRR is about 230~$\mu$m, corresponding to a free spectral range of about 100~GHz. The average Q-factor is about 1.08$\times10^6$. As shown in Fig. \ref{Fig2}\textbf{a}, two continuous-wave (CW) lasers (Pump laser and Aux laser) are amplified by erbium-doped fiber amplifiers (EDFAs) and coupled to the MRR to realize a dual-driven scheme~\cite{zhou2019s,lu2019d,zhang2019sub}. The Aux laser is used to balance the intracavity heat and to facilitate access to the soliton states. By carefully adjusting the frequency detuning between the pump laser and the resonance of the microresonator, the DKS optical comb is generated. The comb has a bandwidth of about 120 nm (from 1480 to 1600 nm), which almost covers the S, C, and L bands (\ref{Fig2}\textbf{b}). Thanks to the dual-driven scheme and the passive isolation of the $Si_3N_4$ microresonator, the soliton state of the comb is stable for a long time, which is essential for long-term QKD applications. See Section \textbf{1} of Supplementary Information for details of the DKS comb. Each frequency line of the DKS optical frequency comb has excellent coherence and can be considered as a single CW laser in our experiment. We select three frequency lines as four-user light sources from the comb via three WDMs, corresponding to the International Telecommunication Union (ITU) 100~GHz DWDM grid at channel 35 (CH35), channel 43 (CH43), and channel 59 (CH59). Then, three 10~GHz bandwidth fiber Bragg grating (FBG) filters are used to clean the frequency spectra of the comb lines (Fig. \ref{Fig2}\textbf{e}-\textbf{g}). A 1$\times$4 coupler divides each frequency line into four portions, and MUXs combine light of different wavelengths into one fiber. Several long fiber spools (2.5~km, 5~km and 7.5~km) are used to introduce random global phase and eliminate single-photon interference (\ref{Fig2}\textbf{d}).  

We use time-bin qubits to encode the bit information via modulated weak coherent pulses (Fig. \ref{Fig2}\textbf{d}). Time-bin qubits are immune to random polarization rotations in fibers, making them suitable for fiber-based quantum communication. In Pauli Z basis, time-bin qubits are encoded as early, $\ket{e}$ or late, $\ket{l}$ for bit values of 0 or 1. In Pauli X basis, the bit values are encoded in their coherent superpositions, $\ket{+}=(\ket{e}+\ket{l})/\sqrt{2}$ and $\ket{-}=(\ket{e}-\ket{l})/\sqrt{2}$, representing bit values of 0 and 1, respectively. The encoder generates the time-bin qubits with a pulse duration of about 0.8 ns, and the time separation between $\ket{e}$ and $\ket{l}$ is about 10 ns. The Z-basis qubits are used for key exchange and the X-basis qubits are used for parameter estimation. In each encoder, the first intensity modulator, IM1, carves the CW light from the DKS comb into a series of narrow pulses, and the second intensity modulator, IM2, implements the decoy-state protocol. The four-intensity decoy-state protocol~\cite{Zhou2016,Jiang2021} has four different intensity states, one signal intensity $z$ in Z basis and three decoy intensities $y$, $x$, $o$ in X basis. A phase modulator (PM) applies a 0 or $\pi$ phase to the later time bin for encoding the X-basis states. In this work, we load the random intensity electrical pulses sequence generated from the arbitrary waveform generator (AWG) on the modulators to encode information. The following VOAs are used to attenuate the pulses to the single-photon level and simulate channel loss.

\begin{figure*}[htbp]
\begin{center}
    \includegraphics[width=0.8\textwidth]{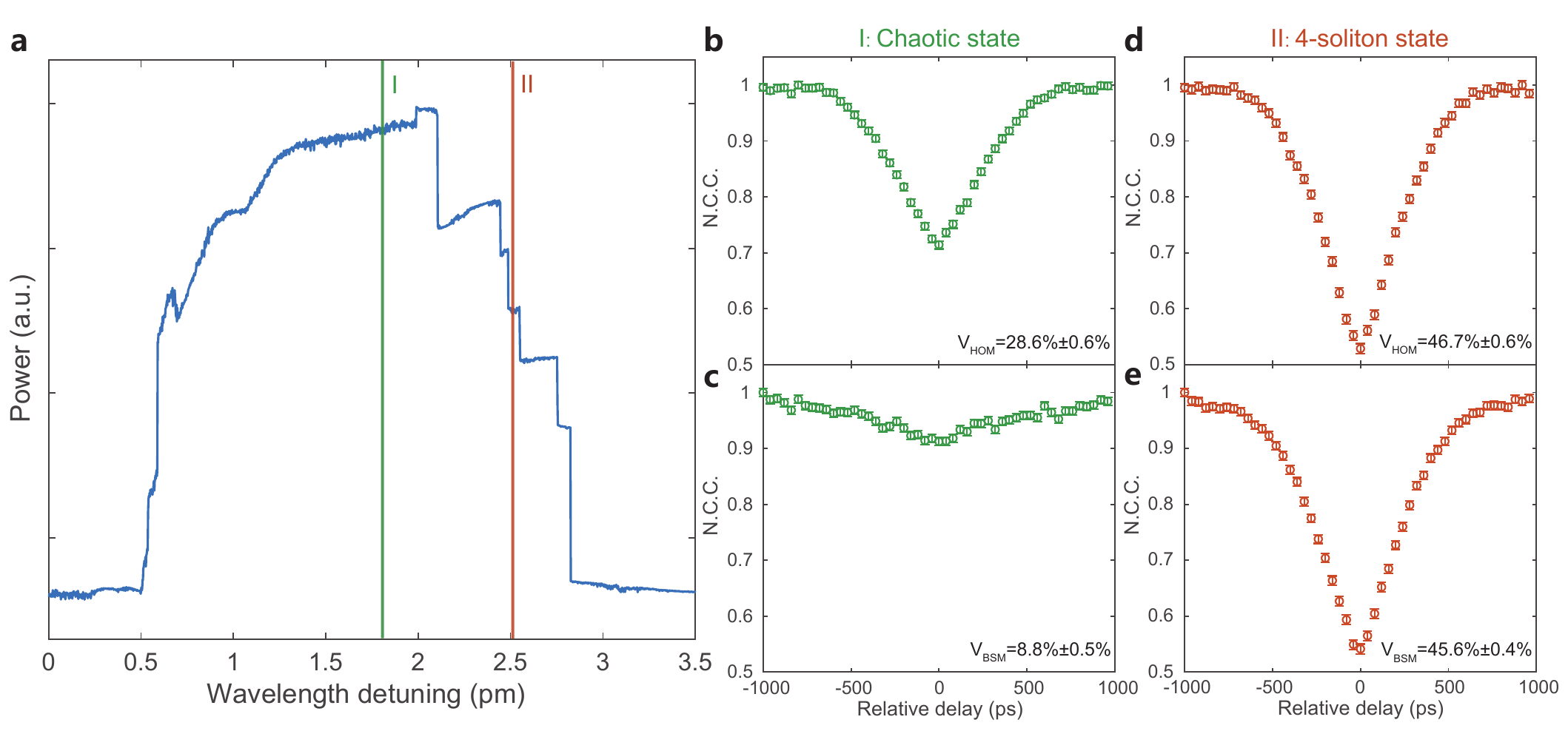}
    \caption{\label{Fig3} Soliton steps of DKS comb and the corresponding HOM interferences and Bell-state measurements with different comb states.  \textbf{a}, The power change of the frequency comb with wavelength detuning between the pump laser and the resonance of MRR, showing the evolution of the soliton steps of DKS comb. \textbf{b}, \textbf{c}, The results of HOM interference (blue circles) and BSM (red circles) with chaotic state of the comb generation, indicated by the green vertical line (marked with $\textup{\uppercase\expandafter{\romannumeral1}}$) in \textbf{a}. \textbf{d}, \textbf{e}, The results of HOM interference (blue circles) and BSM (red circles) with 4-soliton state of the comb generation, indicated by the red vertical line (marked with $\textup{\uppercase\expandafter{\romannumeral2}}$) in \textbf{a}. The excellent coherence of the DKS comb enables the high-visibility interference in HOM and BSM.}
\end{center}
\end{figure*}

At the four-user untrusted relay server (Fig. \ref{Fig2}\textbf{h}), the encoded qubits are routed to six BSM modules depending on their wavelength. Thus, each pair of users implements MDI-QKD simultaneously and independently. The encoded qubits are projected onto the Bell state, $\ket{\Psi^-}=(\ket{el}-\ket{le})/\sqrt{2}$, using a beam splitter (BS) and two superconducting nanowire single-photon detectors (SNSPDs). A field-programmable gate array-based coincidence logic unit with a temporal resolution of about 156 ps is used to record and analyze the photon detection signals.

Six communication links between four users are identified as AB, AC, AD, BC, BD, and CD, respectively. To achieve high-quality BSM, qubits from different users must be indistinguishable in all degrees of freedom (DOF). For temporal DOF, three pairs of users (AB, AC, and AD) are aligned by tuning the relative delays between the electrical signals from the AWG channels. The time differences of the other three pairs (BC, BD, and CD) are compensated by precise optical fiber splicing. Polarization controllers (PCs) are used to eliminate the polarization distinguishability of photons at BSM modules. We investigate the indistinguishability of photons from different users by observing the high visibility of Hong-Ou-Mandel (HOM) interference. The visibilities of all each pair of users are above 46.0\% (see Section \textbf{2} of Supplementary Information for details).

After obtaining the highly indistinguishable time-bin encoded qubits, the next step is to perform BSM on them to realize MDI-QKD. The coincidence counts between two different detectors in different time bins correspond to coincidence counts between $\ket{e}_{D1}$ ($D1$ detects a photon in an early bin) and $\ket{l}_{D2}$ ($D2$ detects a photon in a late bin), or coincidence counts between $\ket{l}_{D1}$ and $\ket{e}_{D2}$. Such a coincidence detection projects two photons onto $\ket{\Psi^-}$ to realize the BSM. In this process, the coherence of the light source is crucial, since a successful BSM relies on the coherent superpositions between the early and late time bins. The DKS optical comb has excellent coherent properties and is therefore well suited for this task. We show the comb states evolution in Fig. \ref{Fig3}\textbf{a}. In Fig. \ref{Fig3}\textbf{b}-\textbf{e}, we show the HOM interference and Bell-state interference results with different comb states. Before the formation of the soliton comb, in the region of the chaotic state (indicated by green line in Fig. \ref{Fig3}\textbf{a}), we obtain the low-visibility HOM interference ($V_{HOM}=28.6\%\pm0.6\%$) and poor-quality BSM result ($V_{BSM}=8.8\%\pm0.5\%$) shown in Fig. \ref{Fig3}\textbf{b} and 3\textbf{c} due to the multiple frequency components and low coherence. Once the soliton state is reached, the high visibility indicates excellent coherence and shows no dependence on the number of solitons. In Fig. \ref{Fig3}\textbf{d} and 3\textbf{e}, we show the high-quality results of the HOM ($V_{HOM}=46.7\%\pm0.6\%$) and BSM ($V_{BSM}=45.6\%\pm0.4\%$) for the highly coherent 4-soliton state of DKS comb. For the MDI-QKD based on two-photon interference as we performed here, the temporal separations between adjacent bins are about 10 ns and maintain cohenrence within this period is crucial. We now directly measure the linewidth of the single comb line with the delayed self-heterodyne method. In Fig. \ref{Fig14}, we obtain 1.4 MHz and 0.11 MHz for the chaotic state and soliton state comb lines, indicating excellent coherent property of soliton combs (see Section \textbf{6} of Supplementary Information for details). 

\begin{figure*}[htbp]
\begin{center}
    \includegraphics[width=0.8\textwidth]{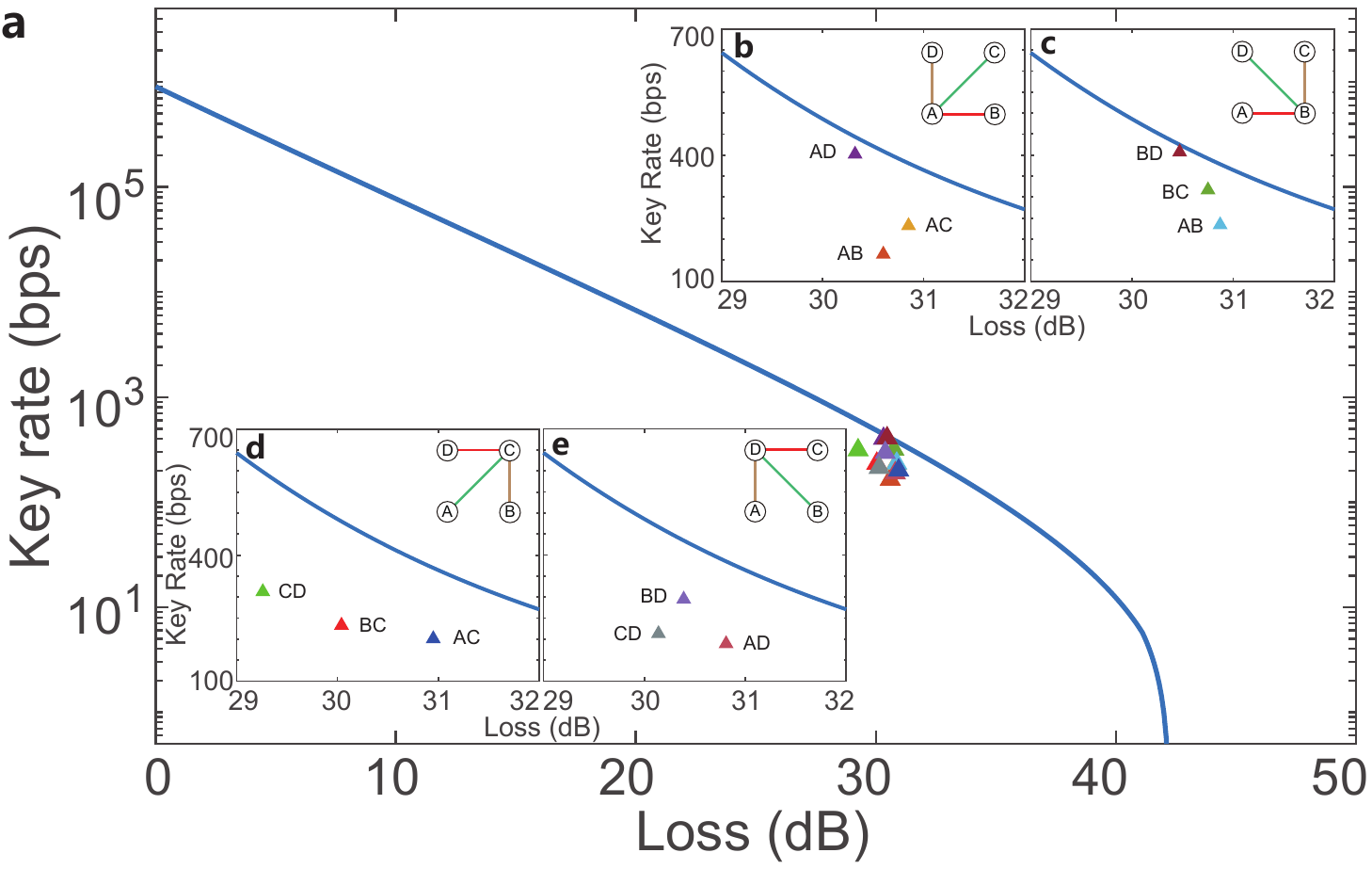}
    \caption{\label{Fig4} The secure key rates with different losses between each user in the four-user MDI-QN. The solid line shows theoretical simulations and the triangular symbols show experimental results of different user pairs with a total transmission loss of about 30 dB. The top-right (\textbf{b}, \textbf{c}) and bottom-left (\textbf{d}, \textbf{e}) insets are the zoom-in views of results for different user pairs with a linear scale.}
\end{center}
\end{figure*}

Then, we implement a four-intensity decoy-state MDI-QKD protocol among four users simultaneously in the fully connected MDI-QN. The finite-key analysis with the Chernoff bound~\cite{curty2014finite,Yin2016,Zhang2017,Jiang2021} is used. The four-intensity parameters are optimized by exploiting the joint constraints on statistical fluctuations with a failure probability of $10^{-10}$. For the total channel loss of 30 dB, the intensities of four states are $z$ = 0.636, $y$ = 0.204, $x$ = 0.054, and $o$ = 0, and the corresponding probabilities of four states are $p_z$ = 0.754, $p_y$ = 0.036, $p_x$ = 0.188, and $p_o$ = 0.022, respectively. Details of the secure key rate analysis are provided in the Section \textbf{3} of Supplementary Information. In our experiment, the MDI-QN operates simultaneously for a time duration of 8.3 hours with a total of $3\times 10^{12}$ sent pulses at a clock rate of 100 MHz for each user pair. After post-processing, the secure key rate results are shown in Fig. \ref{Fig4}. With a total loss of about 30~dB (equivalent to about 150~km of standard single-mode fiber with the propagation loss of  0.2 dB/km), the key rates between different users in our MDI-QN are shown in Fig. \ref{Fig4}\textbf{a}. In Fig. \ref{Fig4}\textbf{b}, we show the zoom-in view of the secure key rate results between user A and the rest of the users in the network, users B, C, D. Similar results between users B/C/D and the rest of the users are shown in Fig. \ref{Fig4}\textbf{c}/\textbf{d}/\textbf{e}. Note that we obtain these results without using any active optical switches to change the network topology, demonstrating the advantages of a fully-connected QKD network. The average user pair key rate in the MDI-QN is 267 bps, which is about three orders of magnitude higher than previous fully connected QN demonstrations~\cite{wengerowsky2018,joshi2020,liu202240,Wen2022}.

\begin{figure*}[htbp]
\begin{center}
    \includegraphics[width=0.8\textwidth]{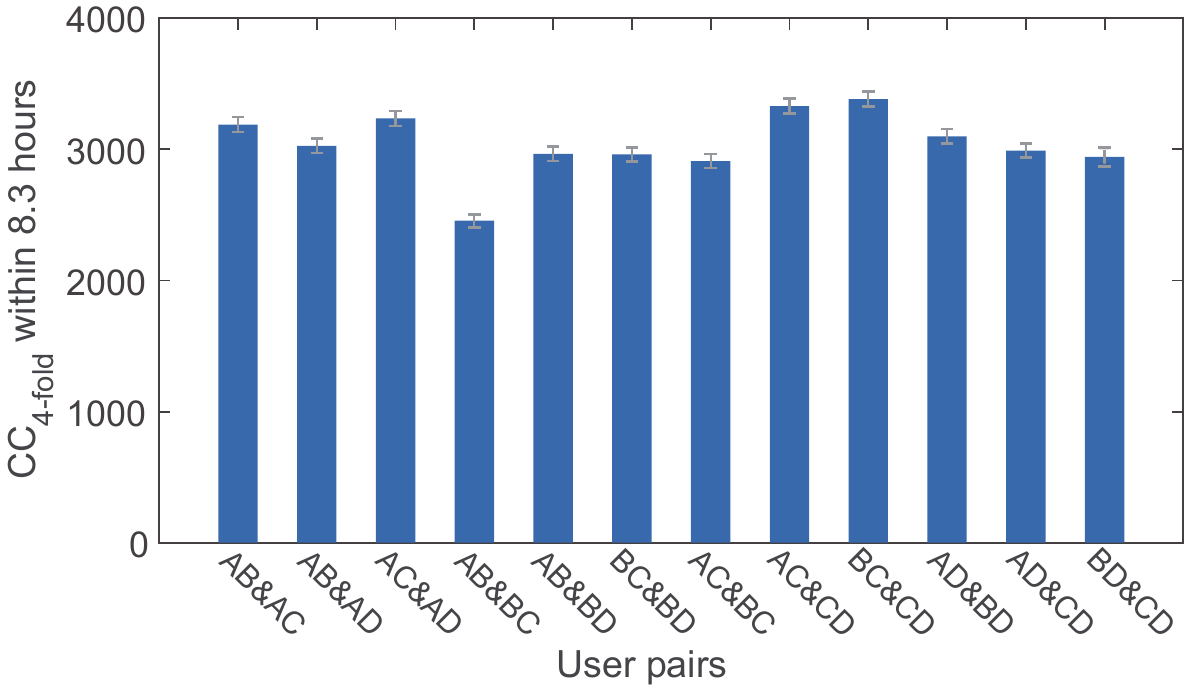}
   \caption{\label{Fig5} Simultaneous communication between two pairs of users in the MDI-QN. The 4-fold coincidence counts (CC) of different user pairs at the total loss of about 30~dB. }
\end{center}
\end{figure*}

More importantly, our fully connected QN has the capability of simultaneous communication between different users, i.e., multiple pairs of users can communicate at the same time. To achieve this goal, we extract four-fold coincidence counts from the raw key data for six pairs of users, representing that two pairs of users (AB$\&$AC, AB$\&$AD, AB$\&$BC, $\ldots$, BD$\&$CD) simultaneously obtain successful $\ket{\Psi^{-}}$ projections within the coincidence time window (1 ns). As shown in Fig. \ref{Fig5}, the average four-fold coincidence counts (CC$_{4-fold}$) are 3033 in 8.3 hours at about 30 dB attenuation, corresponding to 0.1 Hz. We emphasize that this type of simultaneous communication between different pairs of users is not possible in the entanglement-based implementation of the fully connected QNs with TDM method~\cite{wengerowsky2018,joshi2020,liu202240,Wen2022}, since in the previous works only one pair of entangled photons was measured at a time, and thus only two-fold coincidence counts were detected.

In this work, we have proposed and demonstrated a fully connected, trusted node-free, four-user measurement-device-independent quantum key distribution network using an integrated frequency comb for the light source. The novelties of our work can be seen by comparing our work with the well-known protocols and existing QKD networks: 1. The security of our work, especially on the detection side, is guaranteed by the MDI protocol, going beyond the traditional multi-user QKD network based on trusted nodes. 2. The system performance of our work is significantly improved, thus making a decisive step towards the realization of a practical multi-user QKD network. We obtain an average secure key rate of 267 bps with about 30 dB of link attenuation, which is about three orders of magnitude higher than previous entanglement-based demonstrations~\cite{wengerowsky2018,joshi2020,liu202240,Wen2022}. Our work not only has higher key rate, but also allows practical application of QKD in a fully-connected network structure, such as real-time voice telephone with one-time pad encryption~\cite{chen2009field}. More importantly, we achieve simultaneous secure key exchange between two different pairs of users with a rate of 0.1 Hz under about 30 dB attenuation per user pair, which is not achievable in the previous entanglement-based fully connected QNs. As to the simultaneous key exchange, we emphasize this is functionality has never been shown. One possible, yet different, exception might be quantum conference key agreement, using multiphoton state as a resource~\cite{proietti2021}. The simultaneous communication between multiple users will significantly improve with the increase of the system clock rate. For example, the simultaneous raw key rate of the two user pairs is on the order of $10^4$ Hz for 1 GHz clock rate and 10 dB channel loss. Furthermore, our work has the advantage of hardware efficiency. As the light source for the users, we employ the dissipative Kerr soliton optical frequency comb, which provides abundant highly coherent light resources spanning a broad bandwidth with only two external lasers, and thus has the advantage of reducing the laser resource from $O(n)$ to $O(1)$.

Despite these important results, there are several aspects that need to be further improved in our system: 1. Employing the independent integrated DKS comb for each user and implementing the frequency locking among individual comb frequency line. Ideally, each user should own an independent comb as the multiple wavelengths light source. One potential difficulty is to make each comb to be identical in frequency domain. To solve this, one may need frequency dissemination and high-precision frequency translation with side-band modulation, or implement Kerr-induced synchronization \cite{moille2023kerr}.  2. Implementing independent secure key exchange between any pair of users by using an independent encoder for each frequency line of each user. 3. Using real fibers to replace VOAs which simulated channel loss and realizing network field test under real-world environment. The first and second point are within reach with current technology, given the rapid development of integrated photonics in recent years. For each user, integrated turnkey soliton combs~\cite{stern2018battery,shen2020integrated}, whose spectrum can cover multiple communication bands, allow a large number of users to access the network. Note that all users will add an encoder, when a new user is added to the network. This is the hardware requirement for a fully-connected network, demanding new technology advancements. Integrated electronic circuits, vastly increase processing power and reduces the size and cost of computers, and thereby enable the rapid development of the Internet and modern computing systems. Analogous to integrated electronic circuits, high-quality photonic integrated circuits will ease the burden on hardware consumption of quantum networks in the near future. Recent progress in the fabrication of wafer-scale, high-quality thin films of LN-on-insulator have made possible to produce high-performance integrated nanophotonic modulators~\cite{wang2018integrated} in a scalable way. We expect that these type of chip-scale, plug-and-play encoders provide excellent solutions for the upgrade and will become an important component in the QKD network. For the untrusted relay server, waveguide-integrated SNSPDs with high integrated density~\cite{haussler2022scaling} has been realized and will greatly reduce the overhead of detection. Given the rapid development of semiconductor APDs~\cite{comandar2015gigahertz,yan2023compact}, one may use the array of them at the server’s site. In our system, the clock rate is 100 MHz and the pulse width is 0.8 ns. To further enhance the performance of system, one can increase the clock rate. The pulse width will be narrower with the increasing of clock rate. The time jitter of the pulse and the dispersion-compensation in fibers need to be considered carefully. The key rate will be further doubled using time-resolved BSM~\cite{Zheng2021}. What’s more, more pairs of simultaneous communications can be achieved by reducing the channel loss and increasing the system clock rate, which can be solved by increasing the measurable maximum count rate of the detectors, and many efforts ~\cite{Munzberg2018,Zhang2019sspd,rambo202116,Beutel2021,Wei2023hi,grunenfelder2023fast} have been made to solve it. For the third point, many previous laboratory and field QKD implementations have provided valuable experience against environmental disturbance to field fibers. For example, high-speed feedback system was developed for compensating polarization drifting~\cite{PhysRevX.6.011024,Chen2021TF,Liu2021tf}. And the reference-frame-independent protocol with polarization-compensation-free method~\cite{fan2022robust} has demonstrated the improvement of network robustness to environmental disturbances. In addition, we also propose a hardware efficient protocol using the time-division multiplexing (TDM) , as detailed in Section \textbf{5} of Supplementary Information. Users can complete the communication with a relatively simple setup and less resource consumption at the expense of reduction of duty cycle of the system. The performance can be enhanced by improving the clock rate and alternative protocols~\cite{lucamarini2018,ma2018phase,Zeng2022,zhu2023experimental,Xie2022,zhou2022ex}.

\section*{Acknowledgements}
This research was supported by the National Key Research and Development Program of China (Grants No. 2022YFE0137000, 2019YFA0308704), Natural Science Foundation of Jiangsu Province (Grants No. BK20240006, BK20233001), the Leading-Edge Technology Program of Jiangsu Natural Science Foundation (Grant No. BK20192001), the Fundamental Research Funds for the Central Universities, and the Innovation Program for Quantum Science and Technology (Grants No. 2021ZD0300700 and 2021ZD0301500), Supported by the Fundamental Research Funds for the Central Universities (Grant No. 2024300324), and the Nanjing University-China Mobile Communications Group Co., Ltd. Joint Institute.\noindent

\noindent
\bibliographystyle{naturemag}
\bibliography{MDI_ref1}

\begin{thebibliography}{10}
\expandafter\ifx\csname url\endcsname\relax
  \def\url#1{\texttt{#1}}\fi
\expandafter\ifx\csname urlprefix\endcsname\relax\def\urlprefix{URL }\fi
\providecommand{\bibinfo}[2]{#2}
\providecommand{\eprint}[2][]{\url{#2}}

\bibitem{BB84}
\bibinfo{author}{Bennett, C.~H.} \& \bibinfo{author}{Brassard, G.}
\newblock \bibinfo{title}{{Quantum cryptography: Public key distribution and
  coin tossing}}.
\newblock In \emph{\bibinfo{booktitle}{Proceedings of IEEE International
  Conference on Computers, Systems, and Signal Processing}},
  \bibinfo{pages}{175} (\bibinfo{year}{1984}).

\bibitem{Gisin2002}
\bibinfo{author}{Gisin, N.}, \bibinfo{author}{Ribordy, G.},
  \bibinfo{author}{Tittel, W.} \& \bibinfo{author}{Zbinden, H.}
\newblock \bibinfo{title}{Quantum cryptography}.
\newblock \emph{\bibinfo{journal}{Rev. Mod. Phys.}}
  \textbf{\bibinfo{volume}{74}}, \bibinfo{pages}{145--195}
  (\bibinfo{year}{2002}).

\bibitem{Scarani2009}
\bibinfo{author}{Scarani, V.} \emph{et~al.}
\newblock \bibinfo{title}{The security of practical quantum key distribution}.
\newblock \emph{\bibinfo{journal}{Rev. Mod. Phys.}}
  \textbf{\bibinfo{volume}{81}}, \bibinfo{pages}{1301--1350}
  (\bibinfo{year}{2009}).

\bibitem{xu2019}
\bibinfo{author}{Xu, F.}, \bibinfo{author}{Ma, X.}, \bibinfo{author}{Zhang,
  Q.}, \bibinfo{author}{Lo, H.-K.} \& \bibinfo{author}{Pan, J.-W.}
\newblock \bibinfo{title}{Secure quantum key distribution with realistic
  devices}.
\newblock \emph{\bibinfo{journal}{Rev. Mod. Phys.}}
  \textbf{\bibinfo{volume}{92}}, \bibinfo{pages}{025002}
  (\bibinfo{year}{2020}).

\bibitem{Pirandola2020}
\bibinfo{author}{Pirandola, S.} \emph{et~al.}
\newblock \bibinfo{title}{Advances in quantum cryptography}.
\newblock \emph{\bibinfo{journal}{Adv. Opt. Photonics}}
  \textbf{\bibinfo{volume}{12}}, \bibinfo{pages}{1012--1236}
  (\bibinfo{year}{2020}).

\bibitem{PhysRevLett.67.661}
\bibinfo{author}{Ekert, A.~K.}
\newblock \bibinfo{title}{Quantum cryptography based on bell's theorem}.
\newblock \emph{\bibinfo{journal}{Phys. Rev. Lett.}}
  \textbf{\bibinfo{volume}{67}}, \bibinfo{pages}{661--663}
  (\bibinfo{year}{1991}).
\newblock \urlprefix\url{https://link.aps.org/doi/10.1103/PhysRevLett.67.661}.

\bibitem{Brassard2000}
\bibinfo{author}{Brassard, G.}, \bibinfo{author}{L{\"u}tkenhaus, N.},
  \bibinfo{author}{Mor, T.} \& \bibinfo{author}{Sanders, B.~C.}
\newblock \bibinfo{title}{Limitations on practical quantum cryptography}.
\newblock \emph{\bibinfo{journal}{Phys. Rev. Lett.}}
  \textbf{\bibinfo{volume}{85}}, \bibinfo{pages}{1330} (\bibinfo{year}{2000}).

\bibitem{PhysRevA.75.032314}
\bibinfo{author}{Fung, C.-H.~F.}, \bibinfo{author}{Qi, B.},
  \bibinfo{author}{Tamaki, K.} \& \bibinfo{author}{Lo, H.-K.}
\newblock \bibinfo{title}{Phase-remapping attack in practical
  quantum-key-distribution systems}.
\newblock \emph{\bibinfo{journal}{Phys. Rev. A}} \textbf{\bibinfo{volume}{75}},
  \bibinfo{pages}{032314} (\bibinfo{year}{2007}).

\bibitem{Xu_2010}
\bibinfo{author}{Xu, F.}, \bibinfo{author}{Qi, B.} \& \bibinfo{author}{Lo,
  H.-K.}
\newblock \bibinfo{title}{Experimental demonstration of phase-remapping attack
  in a practical quantum key distribution system}.
\newblock \emph{\bibinfo{journal}{N. J. Phys.}} \textbf{\bibinfo{volume}{12}},
  \bibinfo{pages}{113026} (\bibinfo{year}{2010}).

\bibitem{Tang2013}
\bibinfo{author}{Tang, Y.-L.} \emph{et~al.}
\newblock \bibinfo{title}{Source attack of decoy-state quantum key distribution
  using phase information}.
\newblock \emph{\bibinfo{journal}{Phys. Rev. A}} \textbf{\bibinfo{volume}{88}},
  \bibinfo{pages}{022308} (\bibinfo{year}{2013}).

\bibitem{Makarov2006}
\bibinfo{author}{Makarov, V.}, \bibinfo{author}{Anisimov, A.} \&
  \bibinfo{author}{Skaar, J.}
\newblock \bibinfo{title}{Effects of detector efficiency mismatch on security
  of quantum cryptosystems}.
\newblock \emph{\bibinfo{journal}{Phys. Rev. A}} \textbf{\bibinfo{volume}{74}},
  \bibinfo{pages}{022313} (\bibinfo{year}{2006}).

\bibitem{PhysRevA.61.052304}
\bibinfo{author}{L{\"u}tkenhaus, N.}
\newblock \bibinfo{title}{Security against individual attacks for realistic
  quantum key distribution}.
\newblock \emph{\bibinfo{journal}{Phys. Rev. A}} \textbf{\bibinfo{volume}{61}},
  \bibinfo{pages}{052304} (\bibinfo{year}{2000}).

\bibitem{zhao2008}
\bibinfo{author}{Zhao, Y.}, \bibinfo{author}{Fung, C.-H.~F.},
  \bibinfo{author}{Qi, B.}, \bibinfo{author}{Chen, C.} \& \bibinfo{author}{Lo,
  H.-K.}
\newblock \bibinfo{title}{Quantum hacking: Experimental demonstration of
  time-shift attack against practical quantum-key-distribution systems}.
\newblock \emph{\bibinfo{journal}{Phys. Rev. A}} \textbf{\bibinfo{volume}{78}},
  \bibinfo{pages}{042333} (\bibinfo{year}{2008}).

\bibitem{gerhardt2011full}
\bibinfo{author}{Gerhardt, I.} \emph{et~al.}
\newblock \bibinfo{title}{Full-field implementation of a perfect eavesdropper
  on a quantum cryptography system}.
\newblock \emph{\bibinfo{journal}{Nat. Commun.}} \textbf{\bibinfo{volume}{2}},
  \bibinfo{pages}{1--6} (\bibinfo{year}{2011}).

\bibitem{Lydersen_2011}
\bibinfo{author}{Lydersen, L.}, \bibinfo{author}{Akhlaghi, M.~K.},
  \bibinfo{author}{Majedi, A.~H.}, \bibinfo{author}{Skaar, J.} \&
  \bibinfo{author}{Makarov, V.}
\newblock \bibinfo{title}{Controlling a superconducting nanowire single-photon
  detector using tailored bright illumination}.
\newblock \emph{\bibinfo{journal}{N. J. Phys.}} \textbf{\bibinfo{volume}{13}},
  \bibinfo{pages}{113042} (\bibinfo{year}{2011}).

\bibitem{Weier_2011}
\bibinfo{author}{Weier, H.} \emph{et~al.}
\newblock \bibinfo{title}{Quantum eavesdropping without interception: an attack
  exploiting the dead time of single-photon detectors}.
\newblock \emph{\bibinfo{journal}{N. J. Phys.}} \textbf{\bibinfo{volume}{13}},
  \bibinfo{pages}{073024} (\bibinfo{year}{2011}).

\bibitem{braunstein2012side}
\bibinfo{author}{Braunstein, S.~L.} \& \bibinfo{author}{Pirandola, S.}
\newblock \bibinfo{title}{Side-channel-free quantum key distribution}.
\newblock \emph{\bibinfo{journal}{Phys. Rev. Lett.}}
  \textbf{\bibinfo{volume}{108}}, \bibinfo{pages}{130502}
  (\bibinfo{year}{2012}).

\bibitem{Lo2012}
\bibinfo{author}{Lo, H.-K.}, \bibinfo{author}{Curty, M.} \&
  \bibinfo{author}{Qi, B.}
\newblock \bibinfo{title}{Measurement-device-independent quantum key
  distribution}.
\newblock \emph{\bibinfo{journal}{Phys. Rev. Lett.}}
  \textbf{\bibinfo{volume}{108}}, \bibinfo{pages}{130503}
  (\bibinfo{year}{2012}).

\bibitem{ma2012alternative}
\bibinfo{author}{Ma, X.} \& \bibinfo{author}{Razavi, M.}
\newblock \bibinfo{title}{Alternative schemes for
  measurement-device-independent quantum key distribution}.
\newblock \emph{\bibinfo{journal}{Phys. Rev. A}} \textbf{\bibinfo{volume}{86}},
  \bibinfo{pages}{062319} (\bibinfo{year}{2012}).

\bibitem{da2013proof}
\bibinfo{author}{Da~Silva, T.~F.} \emph{et~al.}
\newblock \bibinfo{title}{Proof-of-principle demonstration of
  measurement-device-independent quantum key distribution using polarization
  qubits}.
\newblock \emph{\bibinfo{journal}{Phys. Rev. A}} \textbf{\bibinfo{volume}{88}},
  \bibinfo{pages}{052303} (\bibinfo{year}{2013}).

\bibitem{rubenok2013real}
\bibinfo{author}{Rubenok, A.}, \bibinfo{author}{Slater, J.~A.},
  \bibinfo{author}{Chan, P.}, \bibinfo{author}{Lucio-Martinez, I.} \&
  \bibinfo{author}{Tittel, W.}
\newblock \bibinfo{title}{Real-world two-photon interference and
  proof-of-principle quantum key distribution immune to detector attacks}.
\newblock \emph{\bibinfo{journal}{Phys. Rev. Lett.}}
  \textbf{\bibinfo{volume}{111}}, \bibinfo{pages}{130501}
  (\bibinfo{year}{2013}).

\bibitem{liu2013experimental}
\bibinfo{author}{Liu, Y.} \emph{et~al.}
\newblock \bibinfo{title}{Experimental measurement-device-independent quantum
  key distribution}.
\newblock \emph{\bibinfo{journal}{Phys. Rev. Lett.}}
  \textbf{\bibinfo{volume}{111}}, \bibinfo{pages}{130502}
  (\bibinfo{year}{2013}).

\bibitem{PhysRevLett.122.160501}
\bibinfo{author}{Liu, H.} \emph{et~al.}
\newblock \bibinfo{title}{Experimental demonstration of high-rate
  measurement-device-independent quantum key distribution over asymmetric
  channels}.
\newblock \emph{\bibinfo{journal}{Phys. Rev. Lett.}}
  \textbf{\bibinfo{volume}{122}}, \bibinfo{pages}{160501}
  (\bibinfo{year}{2019}).
\newblock
  \urlprefix\url{https://link.aps.org/doi/10.1103/PhysRevLett.122.160501}.

\bibitem{PhysRevLett.112.190503}
\bibinfo{author}{Tang, Z.} \emph{et~al.}
\newblock \bibinfo{title}{Experimental demonstration of polarization encoding
  measurement-device-independent quantum key distribution}.
\newblock \emph{\bibinfo{journal}{Phys. Rev. Lett.}}
  \textbf{\bibinfo{volume}{112}}, \bibinfo{pages}{190503}
  (\bibinfo{year}{2014}).
\newblock
  \urlprefix\url{https://link.aps.org/doi/10.1103/PhysRevLett.112.190503}.

\bibitem{tang2014}
\bibinfo{author}{Tang, Y.-L.} \emph{et~al.}
\newblock \bibinfo{title}{Measurement-device-independent quantum key
  distribution over 200 km}.
\newblock \emph{\bibinfo{journal}{Phys. Rev. Lett.}}
  \textbf{\bibinfo{volume}{113}}, \bibinfo{pages}{190501}
  (\bibinfo{year}{2014}).

\bibitem{Yin2016}
\bibinfo{author}{Yin, H.-L.} \emph{et~al.}
\newblock \bibinfo{title}{Measurement-device-independent quantum key
  distribution over a 404 km optical fiber}.
\newblock \emph{\bibinfo{journal}{Phys. Rev. Lett.}}
  \textbf{\bibinfo{volume}{117}}, \bibinfo{pages}{190501}
  (\bibinfo{year}{2016}).

\bibitem{comandar2016}
\bibinfo{author}{Comandar, L.} \emph{et~al.}
\newblock \bibinfo{title}{Quantum key distribution without detector
  vulnerabilities using optically seeded lasers}.
\newblock \emph{\bibinfo{journal}{Nat. Photon.}} \textbf{\bibinfo{volume}{10}},
  \bibinfo{pages}{312--315} (\bibinfo{year}{2016}).

\bibitem{Woodward2021}
\bibinfo{author}{Woodward, R.~I.} \emph{et~al.}
\newblock \bibinfo{title}{Gigahertz measurement-device-independent quantum key
  distribution using directly modulated lasers}.
\newblock \emph{\bibinfo{journal}{npj Quantum Inf.}}
  \textbf{\bibinfo{volume}{7}}, \bibinfo{pages}{58} (\bibinfo{year}{2021}).

\bibitem{tang2015}
\bibinfo{author}{Tang, Y.-L.} \emph{et~al.}
\newblock \bibinfo{title}{Field test of measurement-device-independent quantum
  key distribution}.
\newblock \emph{\bibinfo{journal}{IEEE Journal of Selected Topics in Quantum
  Electronics}} \textbf{\bibinfo{volume}{21}}, \bibinfo{pages}{116--122}
  (\bibinfo{year}{2015}).

\bibitem{PhysRevX.6.011024}
\bibinfo{author}{Tang, Y.-L.} \emph{et~al.}
\newblock \bibinfo{title}{Measurement-device-independent quantum key
  distribution over untrustful metropolitan network}.
\newblock \emph{\bibinfo{journal}{Phys. Rev. X}} \textbf{\bibinfo{volume}{6}},
  \bibinfo{pages}{011024} (\bibinfo{year}{2016}).

\bibitem{yuan2020}
\bibinfo{author}{Cao, Y.} \emph{et~al.}
\newblock \bibinfo{title}{Long-distance free-space
  measurement-device-independent quantum key distribution}.
\newblock \emph{\bibinfo{journal}{Phys. Rev. Lett.}}
  \textbf{\bibinfo{volume}{125}}, \bibinfo{pages}{260503}
  (\bibinfo{year}{2020}).

\bibitem{lucamarini2018}
\bibinfo{author}{Lucamarini, M.}, \bibinfo{author}{Yuan, Z.~L.},
  \bibinfo{author}{Dynes, J.~F.} \& \bibinfo{author}{Shields, A.~J.}
\newblock \bibinfo{title}{Overcoming the rate--distance limit of quantum key
  distribution without quantum repeaters}.
\newblock \emph{\bibinfo{journal}{Nature}} \textbf{\bibinfo{volume}{557}},
  \bibinfo{pages}{400--403} (\bibinfo{year}{2018}).

\bibitem{Chen2021TF}
\bibinfo{author}{Chen, J.-P.} \emph{et~al.}
\newblock \bibinfo{title}{Twin-field quantum key distribution over a 511 km
  optical fibre linking two distant metropolitan areas}.
\newblock \emph{\bibinfo{journal}{Nat. Photon.}} \textbf{\bibinfo{volume}{15}},
  \bibinfo{pages}{570--575} (\bibinfo{year}{2021}).

\bibitem{Pittaluga2021}
\bibinfo{author}{Pittaluga, M.} \emph{et~al.}
\newblock \bibinfo{title}{600-km repeater-like quantum communications with
  dual-band stabilization}.
\newblock \emph{\bibinfo{journal}{Nat. Photon.}} \textbf{\bibinfo{volume}{15}},
  \bibinfo{pages}{530--535} (\bibinfo{year}{2021}).

\bibitem{Wang2022}
\bibinfo{author}{Wang, S.} \emph{et~al.}
\newblock \bibinfo{title}{Twin-field quantum key distribution over 830-km
  fibre}.
\newblock \emph{\bibinfo{journal}{Nat. Photon.}} \textbf{\bibinfo{volume}{16}},
  \bibinfo{pages}{154--161} (\bibinfo{year}{2022}).

\bibitem{Liu2021tf}
\bibinfo{author}{Liu, H.} \emph{et~al.}
\newblock \bibinfo{title}{Field test of twin-field quantum key distribution
  through sending-or-not-sending over 428 km}.
\newblock \emph{\bibinfo{journal}{Phys. Rev. Lett.}}
  \textbf{\bibinfo{volume}{126}}, \bibinfo{pages}{250502}
  (\bibinfo{year}{2021}).

\bibitem{li2022twin}
\bibinfo{author}{Li, W.} \emph{et~al.}
\newblock \bibinfo{title}{Twin-field quantum key distribution without phase
  locking}.
\newblock \emph{\bibinfo{journal}{Phys. Rev. Lett.}}
  \textbf{\bibinfo{volume}{130}}, \bibinfo{pages}{250802}
  (\bibinfo{year}{2023}).

\bibitem{zhou2023twin}
\bibinfo{author}{Zhou, L.}, \bibinfo{author}{Lin, J.}, \bibinfo{author}{Jing,
  Y.} \& \bibinfo{author}{Yuan, Z.}
\newblock \bibinfo{title}{Twin-field quantum key distribution without optical
  frequency dissemination}.
\newblock \emph{\bibinfo{journal}{Nat. Commun.}} \textbf{\bibinfo{volume}{14}},
  \bibinfo{pages}{1--8} (\bibinfo{year}{2023}).

\bibitem{liu2023e}
\bibinfo{author}{Liu, Y.} \emph{et~al.}
\newblock \bibinfo{title}{Experimental twin-field quantum key distribution over
  1000 km fiber distance}.
\newblock \emph{\bibinfo{journal}{Phys. Rev. Lett.}}
  \textbf{\bibinfo{volume}{130}}, \bibinfo{pages}{210801}
  (\bibinfo{year}{2023}).

\bibitem{PhysRevLett.124.070501}
\bibinfo{author}{Chen, J.-P.} \emph{et~al.}
\newblock \bibinfo{title}{Sending-or-not-sending with independent lasers:
  Secure twin-field quantum key distribution over 509 km}.
\newblock \emph{\bibinfo{journal}{Phys. Rev. Lett.}}
  \textbf{\bibinfo{volume}{124}}, \bibinfo{pages}{070501}
  (\bibinfo{year}{2020}).
\newblock
  \urlprefix\url{https://link.aps.org/doi/10.1103/PhysRevLett.124.070501}.

\bibitem{Wei2020}
\bibinfo{author}{Wei, K.} \emph{et~al.}
\newblock \bibinfo{title}{High-speed measurement-device-independent quantum key
  distribution with integrated silicon photonics}.
\newblock \emph{\bibinfo{journal}{Phys. Rev. X}} \textbf{\bibinfo{volume}{10}},
  \bibinfo{pages}{031030} (\bibinfo{year}{2020}).

\bibitem{Semenenko2020}
\bibinfo{author}{Semenenko, H.} \emph{et~al.}
\newblock \bibinfo{title}{Chip-based measurement-device-independent quantum key
  distribution}.
\newblock \emph{\bibinfo{journal}{Optica}} \textbf{\bibinfo{volume}{7}},
  \bibinfo{pages}{238--242} (\bibinfo{year}{2020}).

\bibitem{cao2020chip}
\bibinfo{author}{Cao, L.} \emph{et~al.}
\newblock \bibinfo{title}{Chip-based measurement-device-independent quantum key
  distribution using integrated silicon photonic systems}.
\newblock \emph{\bibinfo{journal}{Phys. Rev. Appl.}}
  \textbf{\bibinfo{volume}{14}}, \bibinfo{pages}{011001}
  (\bibinfo{year}{2020}).

\bibitem{Zheng2021}
\bibinfo{author}{Zheng, X.} \emph{et~al.}
\newblock \bibinfo{title}{Heterogeneously integrated, superconducting
  silicon-photonic platform for measurement-device-independent quantum key
  distribution}.
\newblock \emph{\bibinfo{journal}{Adv. Photonics}}
  \textbf{\bibinfo{volume}{3}}, \bibinfo{pages}{055002--055002}
  (\bibinfo{year}{2021}).

\bibitem{Elliott_2002}
\bibinfo{author}{Elliott, C.}
\newblock \bibinfo{title}{Building the quantum network}.
\newblock \emph{\bibinfo{journal}{N. J. Phys.}} \textbf{\bibinfo{volume}{4}},
  \bibinfo{pages}{46--46} (\bibinfo{year}{2002}).

\bibitem{Peev_2009}
\bibinfo{author}{Peev, M.} \emph{et~al.}
\newblock \bibinfo{title}{The {SECOQC} quantum key distribution network in
  {Vienna}}.
\newblock \emph{\bibinfo{journal}{N. J. Phys.}} \textbf{\bibinfo{volume}{11}},
  \bibinfo{pages}{075001} (\bibinfo{year}{2009}).

\bibitem{chen2009field}
\bibinfo{author}{Chen, T.-Y.} \emph{et~al.}
\newblock \bibinfo{title}{Field test of a practical secure communication
  network with decoy-state quantum cryptography}.
\newblock \emph{\bibinfo{journal}{Opt. Express}} \textbf{\bibinfo{volume}{17}},
  \bibinfo{pages}{6540--6549} (\bibinfo{year}{2009}).

\bibitem{sasaki2011field}
\bibinfo{author}{Sasaki, M.} \emph{et~al.}
\newblock \bibinfo{title}{Field test of quantum key distribution in the tokyo
  qkd network}.
\newblock \emph{\bibinfo{journal}{Opt. Express}} \textbf{\bibinfo{volume}{19}},
  \bibinfo{pages}{10387--10409} (\bibinfo{year}{2011}).

\bibitem{chen2021}
\bibinfo{author}{Chen, Y.-A.} \emph{et~al.}
\newblock \bibinfo{title}{An integrated space-to-ground quantum communication
  network over 4,600 kilometres}.
\newblock \emph{\bibinfo{journal}{Nature}} \textbf{\bibinfo{volume}{589}},
  \bibinfo{pages}{214--219} (\bibinfo{year}{2021}).

\bibitem{proietti2021}
\bibinfo{author}{Proietti, M.} \emph{et~al.}
\newblock \bibinfo{title}{Experimental quantum conference key agreement}.
\newblock \emph{\bibinfo{journal}{Sci. Adv.}} \textbf{\bibinfo{volume}{7}},
  \bibinfo{pages}{eabe0395} (\bibinfo{year}{2021}).

\bibitem{huang2023fully}
\bibinfo{author}{Huang, Y.} \emph{et~al.}
\newblock \bibinfo{title}{A fully-connected three-user quantum hyperentangled
  network}.
\newblock \emph{\bibinfo{journal}{Quantum Front.}}
  \textbf{\bibinfo{volume}{2}}, \bibinfo{pages}{4} (\bibinfo{year}{2023}).

\bibitem{frohlich2013quantum}
\bibinfo{author}{Fr{\"o}hlich, B.} \emph{et~al.}
\newblock \bibinfo{title}{A quantum access network}.
\newblock \emph{\bibinfo{journal}{Nature}} \textbf{\bibinfo{volume}{501}},
  \bibinfo{pages}{69--72} (\bibinfo{year}{2013}).

\bibitem{wengerowsky2018}
\bibinfo{author}{Wengerowsky, S.}, \bibinfo{author}{Joshi, S.~K.},
  \bibinfo{author}{Steinlechner, F.}, \bibinfo{author}{H{\"u}bel, H.} \&
  \bibinfo{author}{Ursin, R.}
\newblock \bibinfo{title}{An entanglement-based wavelength-multiplexed quantum
  communication network}.
\newblock \emph{\bibinfo{journal}{Nature}} \textbf{\bibinfo{volume}{564}},
  \bibinfo{pages}{225--228} (\bibinfo{year}{2018}).

\bibitem{joshi2020}
\bibinfo{author}{Joshi, S.~K.} \emph{et~al.}
\newblock \bibinfo{title}{A trusted node-free eight-user metropolitan quantum
  communication network}.
\newblock \emph{\bibinfo{journal}{Sci. Adv.}} \textbf{\bibinfo{volume}{6}},
  \bibinfo{pages}{eaba0959} (\bibinfo{year}{2020}).

\bibitem{liu202240}
\bibinfo{author}{Liu, X.} \emph{et~al.}
\newblock \bibinfo{title}{40-user fully connected entanglement-based quantum
  key distribution network without trusted node}.
\newblock \emph{\bibinfo{journal}{PhotoniX}} \textbf{\bibinfo{volume}{3}},
  \bibinfo{pages}{1--15} (\bibinfo{year}{2022}).

\bibitem{Wen2022}
\bibinfo{author}{Wen, W.} \emph{et~al.}
\newblock \bibinfo{title}{Realizing an entanglement-based multiuser quantum
  network with integrated photonics}.
\newblock \emph{\bibinfo{journal}{Phys. Rev. Appl.}}
  \textbf{\bibinfo{volume}{18}}, \bibinfo{pages}{024059}
  (\bibinfo{year}{2022}).

\bibitem{herr2014temporal}
\bibinfo{author}{Herr, T.} \emph{et~al.}
\newblock \bibinfo{title}{Temporal solitons in optical microresonators}.
\newblock \emph{\bibinfo{journal}{Nat. Photon.}} \textbf{\bibinfo{volume}{8}},
  \bibinfo{pages}{145--152} (\bibinfo{year}{2014}).

\bibitem{kippenberg2018d}
\bibinfo{author}{Kippenberg, T.~J.}, \bibinfo{author}{Gaeta, A.~L.},
  \bibinfo{author}{Lipson, M.} \& \bibinfo{author}{Gorodetsky, M.~L.}
\newblock \bibinfo{title}{Dissipative kerr solitons in optical
  microresonators}.
\newblock \emph{\bibinfo{journal}{Science}} \textbf{\bibinfo{volume}{361}},
  \bibinfo{pages}{eaan8083} (\bibinfo{year}{2018}).

\bibitem{wang2020quantum}
\bibinfo{author}{Wang, F.} \emph{et~al.}
\newblock \bibinfo{title}{Quantum key distribution with on-chip dissipative
  kerr soliton}.
\newblock \emph{\bibinfo{journal}{Laser Photonics Rev.}}
  \textbf{\bibinfo{volume}{14}}, \bibinfo{pages}{1900190}
  (\bibinfo{year}{2020}).

\bibitem{Beutel2021}
\bibinfo{author}{Beutel, F.}, \bibinfo{author}{Gehring, H.},
  \bibinfo{author}{Wolff, M.~A.}, \bibinfo{author}{Schuck, C.} \&
  \bibinfo{author}{Pernice, W.}
\newblock \bibinfo{title}{Detector-integrated on-chip qkd receiver for {GHz}
  clock rates}.
\newblock \emph{\bibinfo{journal}{npj Quantum Inf.}}
  \textbf{\bibinfo{volume}{7}}, \bibinfo{pages}{40} (\bibinfo{year}{2021}).

\bibitem{haussler2022scaling}
\bibinfo{author}{Haussler, M.} \emph{et~al.}
\newblock \bibinfo{title}{Scaling waveguide-integrated superconducting nanowire
  single-photon detector solutions to large numbers of independent optical
  channels}.
\newblock \emph{\bibinfo{journal}{Rev. Sci. Instrum.}}
  \textbf{\bibinfo{volume}{94}}, \bibinfo{pages}{013103}
  (\bibinfo{year}{2023}).

\bibitem{zhou2019s}
\bibinfo{author}{Zhou, H.} \emph{et~al.}
\newblock \bibinfo{title}{Soliton bursts and deterministic dissipative kerr
  soliton generation in auxiliary-assisted microcavities}.
\newblock \emph{\bibinfo{journal}{Light: Sci. Appl.}}
  \textbf{\bibinfo{volume}{8}}, \bibinfo{pages}{50} (\bibinfo{year}{2019}).

\bibitem{lu2019d}
\bibinfo{author}{Lu, Z.} \emph{et~al.}
\newblock \bibinfo{title}{Deterministic generation and switching of dissipative
  kerr soliton in a thermally controlled micro-resonator}.
\newblock \emph{\bibinfo{journal}{AIP Adv.}} \textbf{\bibinfo{volume}{9}},
  \bibinfo{pages}{025314} (\bibinfo{year}{2019}).

\bibitem{zhang2019sub}
\bibinfo{author}{Zhang, S.} \emph{et~al.}
\newblock \bibinfo{title}{Sub-milliwatt-level microresonator solitons with
  extended access range using an auxiliary laser}.
\newblock \emph{\bibinfo{journal}{Optica}} \textbf{\bibinfo{volume}{6}},
  \bibinfo{pages}{206--212} (\bibinfo{year}{2019}).

\bibitem{Zhou2016}
\bibinfo{author}{Zhou, Y.-H.}, \bibinfo{author}{Yu, Z.-W.} \&
  \bibinfo{author}{Wang, X.-B.}
\newblock \bibinfo{title}{Making the decoy-state measurement-device-independent
  quantum key distribution practically useful}.
\newblock \emph{\bibinfo{journal}{Phys. Rev. A}} \textbf{\bibinfo{volume}{93}},
  \bibinfo{pages}{042324} (\bibinfo{year}{2016}).

\bibitem{Zhang2017}
\bibinfo{author}{Zhang, Z.}, \bibinfo{author}{Zhao, Q.},
  \bibinfo{author}{Razavi, M.} \& \bibinfo{author}{Ma, X.}
\newblock \bibinfo{title}{Improved key-rate bounds for practical decoy-state
  quantum-key-distribution systems}.
\newblock \emph{\bibinfo{journal}{Phys. Rev. A}} \textbf{\bibinfo{volume}{95}},
  \bibinfo{pages}{012333} (\bibinfo{year}{2017}).

\bibitem{Jiang2021}
\bibinfo{author}{Jiang, C.}, \bibinfo{author}{Yu, Z.-W.}, \bibinfo{author}{Hu,
  X.-L.} \& \bibinfo{author}{Wang, X.-B.}
\newblock \bibinfo{title}{Higher key rate of measurement-device-independent
  quantum key distribution through joint data processing}.
\newblock \emph{\bibinfo{journal}{Phys. Rev. A}}
  \textbf{\bibinfo{volume}{103}}, \bibinfo{pages}{012402}
  (\bibinfo{year}{2021}).

\bibitem{curty2014finite}
\bibinfo{author}{Curty, M.} \emph{et~al.}
\newblock \bibinfo{title}{Finite-key analysis for
  measurement-device-independent quantum key distribution}.
\newblock \emph{\bibinfo{journal}{Nat. Commun.}} \textbf{\bibinfo{volume}{5}},
  \bibinfo{pages}{1--7} (\bibinfo{year}{2014}).

\bibitem{moille2023kerr}
\bibinfo{author}{Moille, G.} \emph{et~al.}
\newblock \bibinfo{title}{Kerr-induced synchronization of a cavity soliton to
  an optical reference}.
\newblock \emph{\bibinfo{journal}{Nature}} \textbf{\bibinfo{volume}{624}},
  \bibinfo{pages}{267--274} (\bibinfo{year}{2023}).

\bibitem{stern2018battery}
\bibinfo{author}{Stern, B.}, \bibinfo{author}{Ji, X.},
  \bibinfo{author}{Okawachi, Y.}, \bibinfo{author}{Gaeta, A.~L.} \&
  \bibinfo{author}{Lipson, M.}
\newblock \bibinfo{title}{Battery-operated integrated frequency comb
  generator}.
\newblock \emph{\bibinfo{journal}{Nature}} \textbf{\bibinfo{volume}{562}},
  \bibinfo{pages}{401--405} (\bibinfo{year}{2018}).

\bibitem{shen2020integrated}
\bibinfo{author}{Shen, B.} \emph{et~al.}
\newblock \bibinfo{title}{Integrated turnkey soliton microcombs}.
\newblock \emph{\bibinfo{journal}{Nature}} \textbf{\bibinfo{volume}{582}},
  \bibinfo{pages}{365--369} (\bibinfo{year}{2020}).

\bibitem{wang2018integrated}
\bibinfo{author}{Wang, C.} \emph{et~al.}
\newblock \bibinfo{title}{Integrated lithium niobate electro-optic modulators
  operating at cmos-compatible voltages}.
\newblock \emph{\bibinfo{journal}{Nature}} \textbf{\bibinfo{volume}{562}},
  \bibinfo{pages}{101--104} (\bibinfo{year}{2018}).

\bibitem{comandar2015gigahertz}
\bibinfo{author}{Comandar, L.~C.} \emph{et~al.}
\newblock \bibinfo{title}{Gigahertz-gated {InGaAs/InP} single-photon detector
  with detection efficiency exceeding 55\% at 1550 nm}.
\newblock \emph{\bibinfo{journal}{J. Appl. Phys.}}
  \textbf{\bibinfo{volume}{117}} (\bibinfo{year}{2015}).

\bibitem{yan2023compact}
\bibinfo{author}{Yan, Z.}, \bibinfo{author}{Shi, T.}, \bibinfo{author}{Fan,
  Y.}, \bibinfo{author}{Zhou, L.} \& \bibinfo{author}{Yuan, Z.}
\newblock \bibinfo{title}{Compact {InGaAs/InP} single-photon detector module
  with ultra-narrowband interference circuits}.
\newblock \emph{\bibinfo{journal}{Adv. Dev. Instrum.}}
  \textbf{\bibinfo{volume}{4}}, \bibinfo{pages}{0029} (\bibinfo{year}{2023}).

\bibitem{Munzberg2018}
\bibinfo{author}{Münzberg, J.} \emph{et~al.}
\newblock \bibinfo{title}{Superconducting nanowire single-photon detector
  implemented in a 2{D} photonic crystal cavity}.
\newblock \emph{\bibinfo{journal}{Optica}} \textbf{\bibinfo{volume}{5}},
  \bibinfo{pages}{658--665} (\bibinfo{year}{2018}).

\bibitem{Zhang2019sspd}
\bibinfo{author}{Zhang, W.} \emph{et~al.}
\newblock \bibinfo{title}{A 16-pixel interleaved superconducting nanowire
  single-photon detector array with a maximum count rate exceeding 1.5 {GH}z}.
\newblock \emph{\bibinfo{journal}{IEEE Trans. Appl. Supercond.}}
  \textbf{\bibinfo{volume}{29}}, \bibinfo{pages}{1--4} (\bibinfo{year}{2019}).

\bibitem{rambo202116}
\bibinfo{author}{Rambo, T.~M.}, \bibinfo{author}{Conover, A.~R.} \&
  \bibinfo{author}{Miller, A.~J.}
\newblock \bibinfo{title}{16-element superconducting nanowire single-photon
  detector for gigahertz counting at 1550-nm}.
\newblock \emph{\bibinfo{journal}{arXiv preprint arXiv:2103.14086}}
  (\bibinfo{year}{2021}).

\bibitem{Wei2023hi}
\bibinfo{author}{Li, W.} \emph{et~al.}
\newblock \bibinfo{title}{High-rate quantum key distribution exceeding
  110~{M}b~s$^{–1}$}.
\newblock \emph{\bibinfo{journal}{Nat. Photon.}} \textbf{\bibinfo{volume}{17}},
  \bibinfo{pages}{416--421} (\bibinfo{year}{2023}).

\bibitem{grunenfelder2023fast}
\bibinfo{author}{Gr{\"u}nenfelder, F.} \emph{et~al.}
\newblock \bibinfo{title}{Fast single-photon detectors and real-time key
  distillation enable high secret-key-rate quantum key distribution systems}.
\newblock \emph{\bibinfo{journal}{Nat. Photon.}} \textbf{\bibinfo{volume}{17}},
  \bibinfo{pages}{422--426} (\bibinfo{year}{2023}).

\bibitem{fan2022robust}
\bibinfo{author}{Fan-Yuan, G.-J.} \emph{et~al.}
\newblock \bibinfo{title}{Robust and adaptable quantum key distribution network
  without trusted nodes}.
\newblock \emph{\bibinfo{journal}{Optica}} \textbf{\bibinfo{volume}{9}},
  \bibinfo{pages}{812--823} (\bibinfo{year}{2022}).

\bibitem{ma2018phase}
\bibinfo{author}{Ma, X.}, \bibinfo{author}{Zeng, P.} \& \bibinfo{author}{Zhou,
  H.}
\newblock \bibinfo{title}{Phase-matching quantum key distribution}.
\newblock \emph{\bibinfo{journal}{Phys. Rev. X}} \textbf{\bibinfo{volume}{8}},
  \bibinfo{pages}{031043} (\bibinfo{year}{2018}).

\bibitem{Zeng2022}
\bibinfo{author}{Zeng, P.}, \bibinfo{author}{Zhou, H.}, \bibinfo{author}{Wu,
  W.} \& \bibinfo{author}{Ma, X.}
\newblock \bibinfo{title}{Mode-pairing quantum key distribution}.
\newblock \emph{\bibinfo{journal}{Nat. Commun.}} \textbf{\bibinfo{volume}{13}},
  \bibinfo{pages}{3903} (\bibinfo{year}{2022}).

\bibitem{zhu2023experimental}
\bibinfo{author}{Zhu, H.~T.} \emph{et~al.}
\newblock \bibinfo{title}{Experimental mode-pairing
  measurement-device-independent quantum key distribution without global phase
  locking}.
\newblock \emph{\bibinfo{journal}{Phys. Rev. Lett.}}
  \textbf{\bibinfo{volume}{130}}, \bibinfo{pages}{030801}
  (\bibinfo{year}{2023}).

\bibitem{Xie2022}
\bibinfo{author}{Xie, Y.-M.} \emph{et~al.}
\newblock \bibinfo{title}{Breaking the rate-loss bound of quantum key
  distribution with asynchronous two-photon interference}.
\newblock \emph{\bibinfo{journal}{PRX Quantum}} \textbf{\bibinfo{volume}{3}},
  \bibinfo{pages}{020315} (\bibinfo{year}{2022}).

\bibitem{zhou2022ex}
\bibinfo{author}{Zhou, L.} \emph{et~al.}
\newblock \bibinfo{title}{Experimental quantum communication overcomes the
  rate-loss limit without global phase tracking}.
\newblock \emph{\bibinfo{journal}{Phys. Rev. Lett.}}
  \textbf{\bibinfo{volume}{130}}, \bibinfo{pages}{250801}
  (\bibinfo{year}{2023}).

\end{thebibliography}

\clearpage
\newpage

\section{Supplementary information}

\subsection{DKS frequency comb source}\label{sec:one}
\begin{figure*}[htbp]
\begin{center}
    \includegraphics[width=0.8\textwidth]{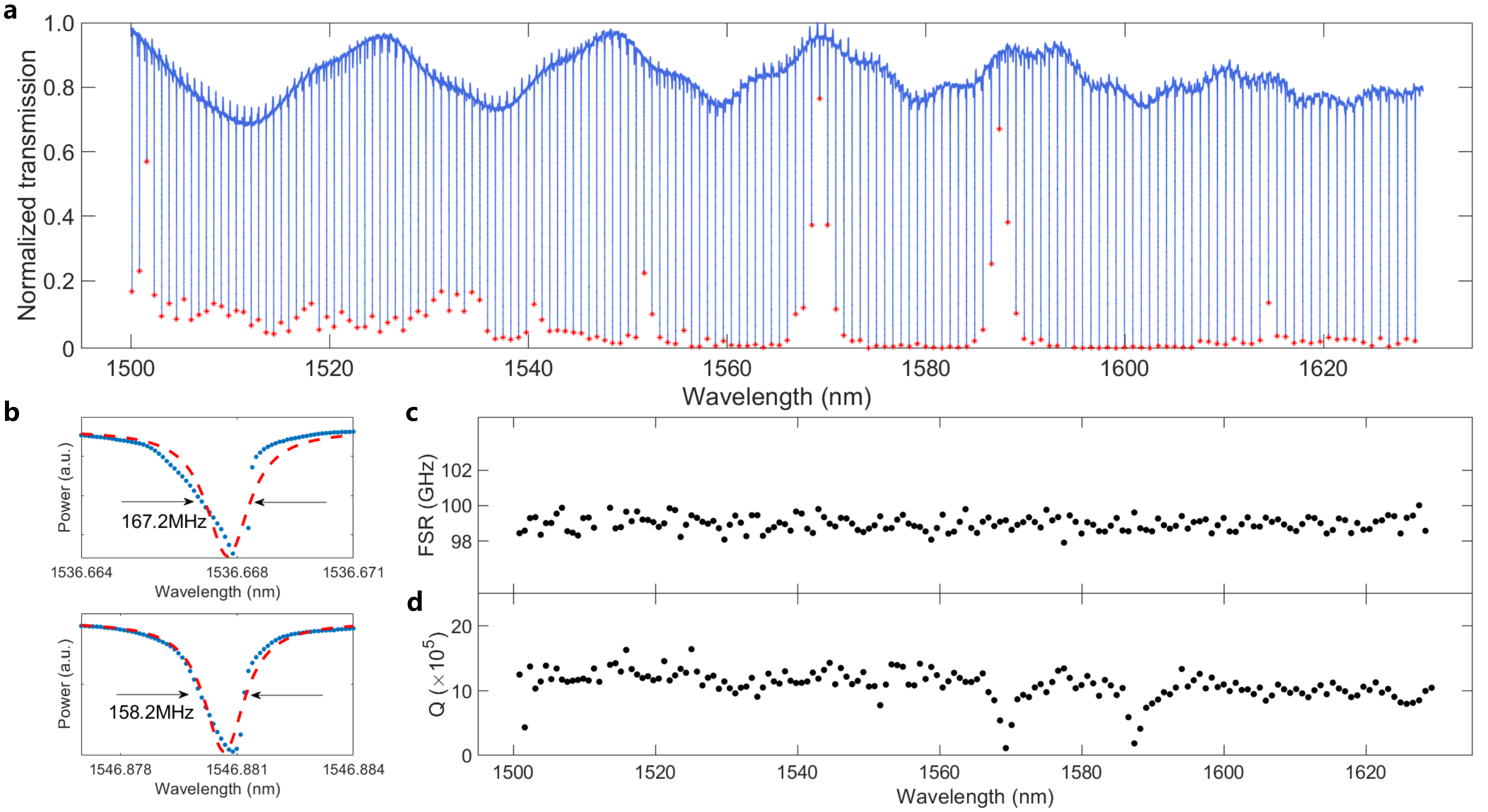}
    \caption{\label{Fig6} Characterization of the Silicon Nitride microring with laser. \textbf{a}, The measured cavity transmission spectrum ranging from 1500 to 1630nm. \textbf{b}, The raw data and fitting line of the pump laser resonance and auxiliary pump laser resonance. \textbf{c} and \textbf{d}, The free spectrum range (FSR) and fitted quality factor (Q) of the resonator.}
\end{center}
\end{figure*}

To characterize the integrated silicon nitride microring resonator (MRR), we measure the transmission spectrum to select two whispering gallery mode wavelengths with similar parameters for dual-driven scheme~\cite{zhou2019s,lu2019d,zhang2019sub}, as shown in Fig. \ref{Fig6}\textbf{a}. Pump laser is fixed at 1536.6678~nm (CH51) with an FWHM of about 167.2~MHz, and the auxiliary laser is fixed at 1546.8807~nm (CH37) with an FWHM of about 158.2~MHz, as shown in Fig. \ref{Fig6}\textbf{b}. The radius of the microring is about 230~$\mu$m, corresponding to a free spectral range (FSR) of about 100~GHz in transverse electric (TE) mode. Light is coupled into and out from the chip by a high numerical aperture fiber array with a pitch of 127 $\mu$m with insertion loss of about 6.58~dB. The MRR is evanescently coupled to a single-bus waveguide via a nominal 500~nm gap point coupler. Both the ring and waveguide have a cross-section of 1.6~$\mu$m wide and 0.8~$\mu$m high. By calculating 122 cavity resonance modes, the averaged FSR is 98.7498~GHz (Fig. \ref{Fig6}\textbf{c}) and the averaged measured quality factor (Q) is about $1.08\times10^6$ (Fig. \ref{Fig6}\textbf{d}).

In order to generate the dissipative Kerr soliton (DKS) optical frequency comb (OFC) in a easy-controlled way and maintain its long-term operation stability, we use the dual-driven scheme~\cite{zhou2019s,lu2019d,zhang2019sub} as shown in the main text. Lights from two CW lasers are coupled into the micro-cavity from opposite directions. One CW laser called pump laser (Santec TSL-770) is amplified (Keoposys EDFA) at 1W in the clockwise direction, and another called Aux laser (Santec TSL-550) at about 1.58~W in the counter-clockwise direction. The light is adjusted to be transverse electric ($TE_{00}$) mode by a fiber polarization controller on both paths. Two cascaded circulators with more than 60~dB isolation are used to prevent mutual damage of two lasers. 

\begin{figure*}[htbp]
\begin{center}
    \includegraphics[width=0.8\textwidth]{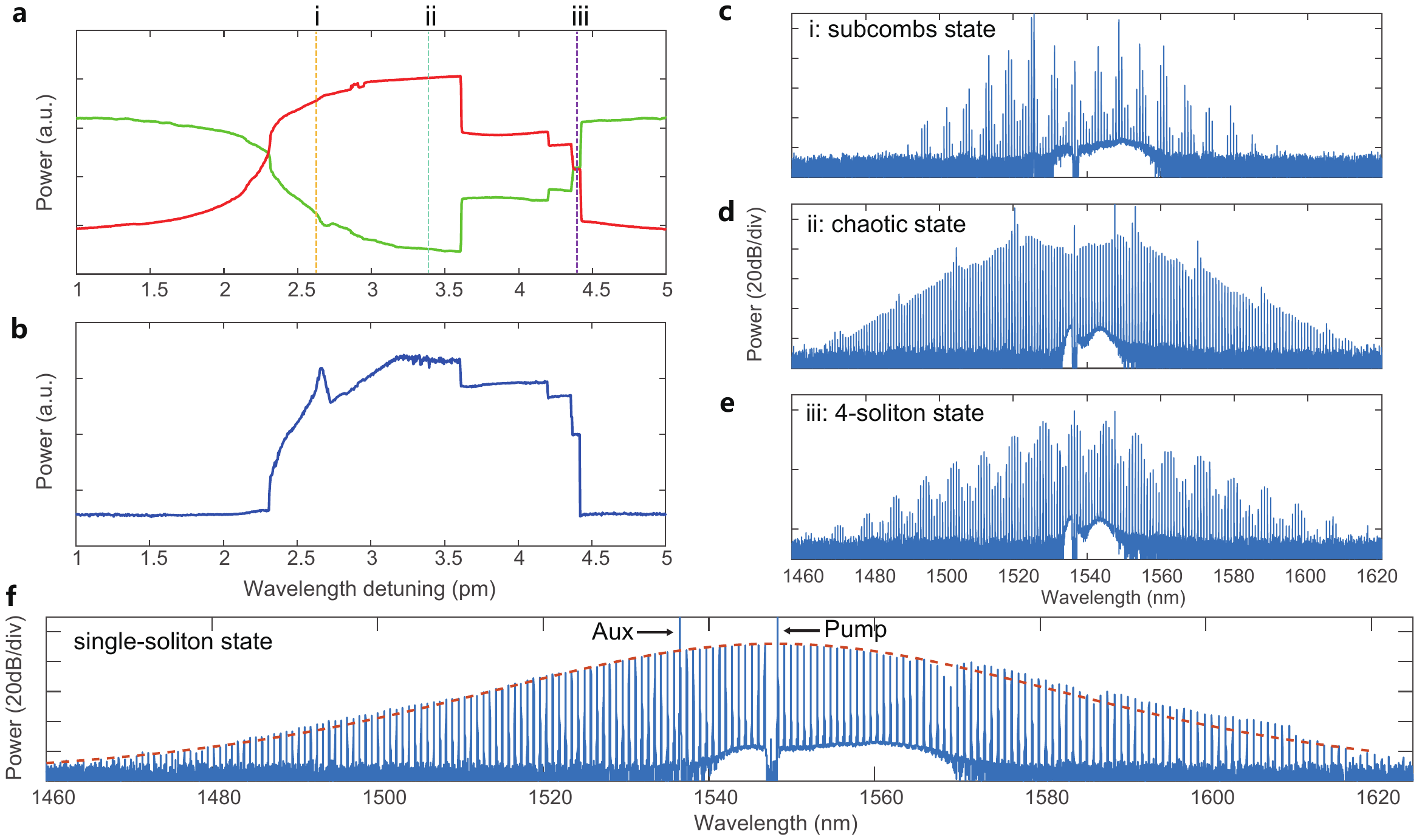}
    \caption{\label{Fig7} The generation of the Kerr frequency comb. \textbf{a}, The measured power evolutions as a 1.0~W pump laser adiabatically scans from the blue- to red-detuning regimes of a cavity mode. The solid green trace correspond to the power of pump laser and red trace correspond to the power of auxiliary laser. \textbf{b}, The solid blue lines correspond to the power of DKS optical frequency comb. \textbf{c}-\textbf{e}, Optical spectral snapshots of the generated DKS optical frequency comb with the low-noise subcombs state (indicated by line i in \textbf{a}), chaotic state (indicated by line ii in \textbf{a}) and 4-soliton state (indicated by line iii in \textbf{a}). \textbf{f}, Optical spectral lines of the DKS optical frequency comb with the single-soliton state. And the red dotted line is the fitting line with a $sech^2$ envelope.}
\end{center}
\end{figure*}

By employing the dual-driven scheme and thus separating the cavity thermal nonlinearities from the Kerr dynamics, we achieve a stable DKS comb. Firstly, an auxiliary laser frequency is turned into a resonance (1546.88~nm) following the traditional self-thermal locking trajectory and is stopped near the resonance peak but still kept blue-detuned. Subsequently, the pump laser is turned into another location(1536.67~nm) in the counter-propagating direction. We observe that when the pump traverses from the blue-detuning to the red-detuning regime (Fig. \ref{Fig7}\textbf{a}-\textbf{b}), stable comb spectra can be formed, as shown in Fig. \ref{Fig7}\textbf{c}-\textbf{f}. Specifically, when the pump is blue-detuned, subcombs with well separated line doublets (i.e., spacing multiple FSR) are generated, as shown in Fig. \ref{Fig7}\textbf{c}. When the pump enters the red-detuned regime, the oscillator achieves gap-free single FSR comb spectra resembling multiple DKS waveforms (Fig. \ref{Fig7}\textbf{d}). Upon further tuning the pump toward the red-sided regime, we observe step-like discrete soliton state jumps,such as 4-soliton state (Fig. \ref{Fig7}\textbf{e}) and eventually, a single-soliton microcomb with a $sech^2$ envelope is achieved via fine-tuning frequency tuning, as shown in Fig. \ref{Fig7}\textbf{f}.

\begin{figure*}[htbp]
\begin{center}
    \includegraphics[width=0.8\textwidth]{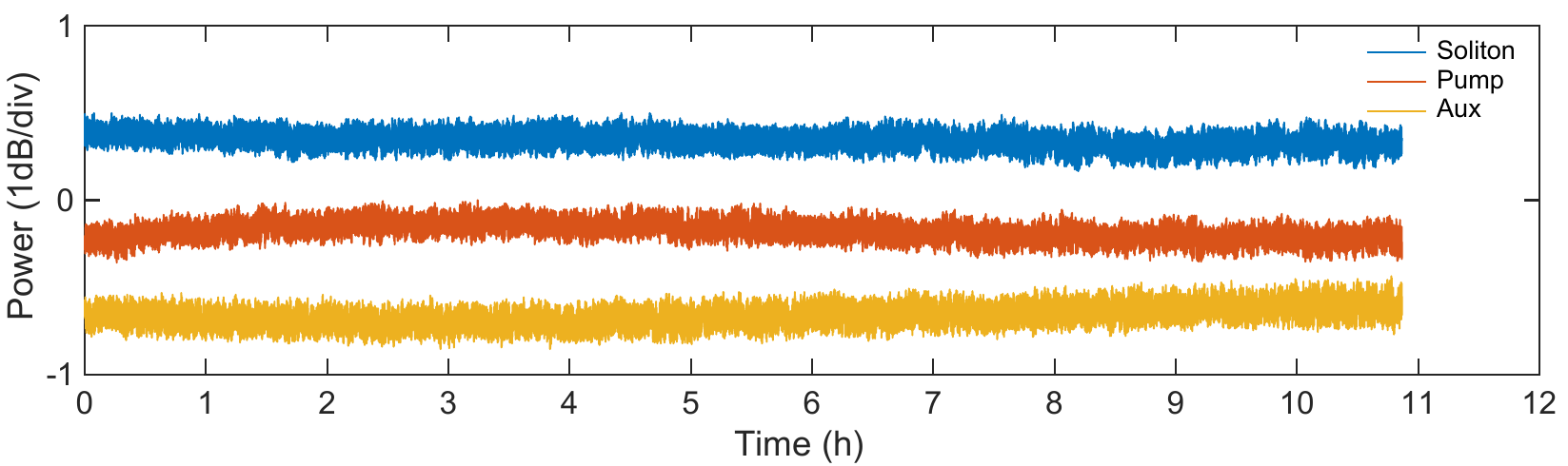}
    \caption{\label{Fig8} The power stability of soliton comb.}
\end{center}
\end{figure*}

At the resonator output, a WDM of CH51 (auxiliary light is at CH37) is used to separate the pump light and DKS comb. The power of the pump laser and auxiliary laser are sent to two power meters (Thorlab PM100D), which is used as monitors of the stability of soliton comb along with spectrum measurements (Fig. \ref{Fig2} in the main text). As shown in Fig. \ref{Fig8}, the soliton state of comb is stable more than 10 hours.

\subsection{Relative delay scanning}\label{sec:two}
\begin{figure*}[htbp]
\begin{center}
    \includegraphics[width=0.8\textwidth]{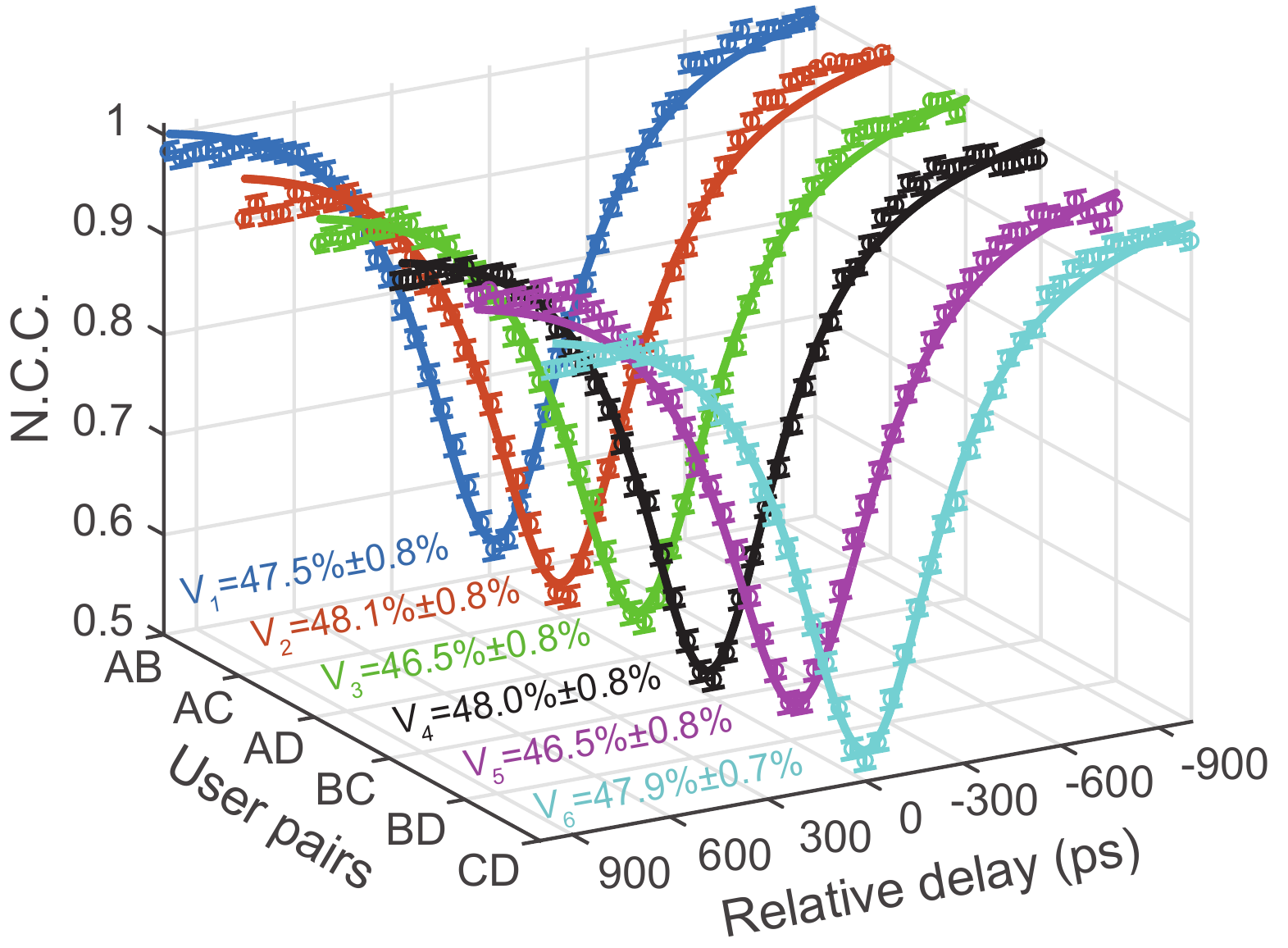}
    \caption{\label{Fig9} Hong-Ou-Mandel (HOM) interference results between different users. The circles are experimental data which correspond to normalized coincidence count of two detectors at different relative delay and the solid curves are fitted curves, see text for details.}
\end{center}
\end{figure*}

We scan the relative electrical delays of AWG channels and obtain coincidence counts of different BSM modules. Hong-Ou-Mandel (HOM) interference results as a function of relative electrical delays between different users are shown in Fig. \ref{Fig9}. HOM interference visibilities of $47.5\% \pm 0.8\%$, $48.1\% \pm 0.8\%$, $46.5\% \pm 0.8\%$, $48.0\% \pm 0.8\%$, $46.5\% \pm 0.8\%$ and $47.9\% \pm 0.7\%$ are observed for AB, AC, AD, BC, BD and CD, respectively.

\begin{figure*}[htbp]
\begin{center}
    \includegraphics[width=1\textwidth]{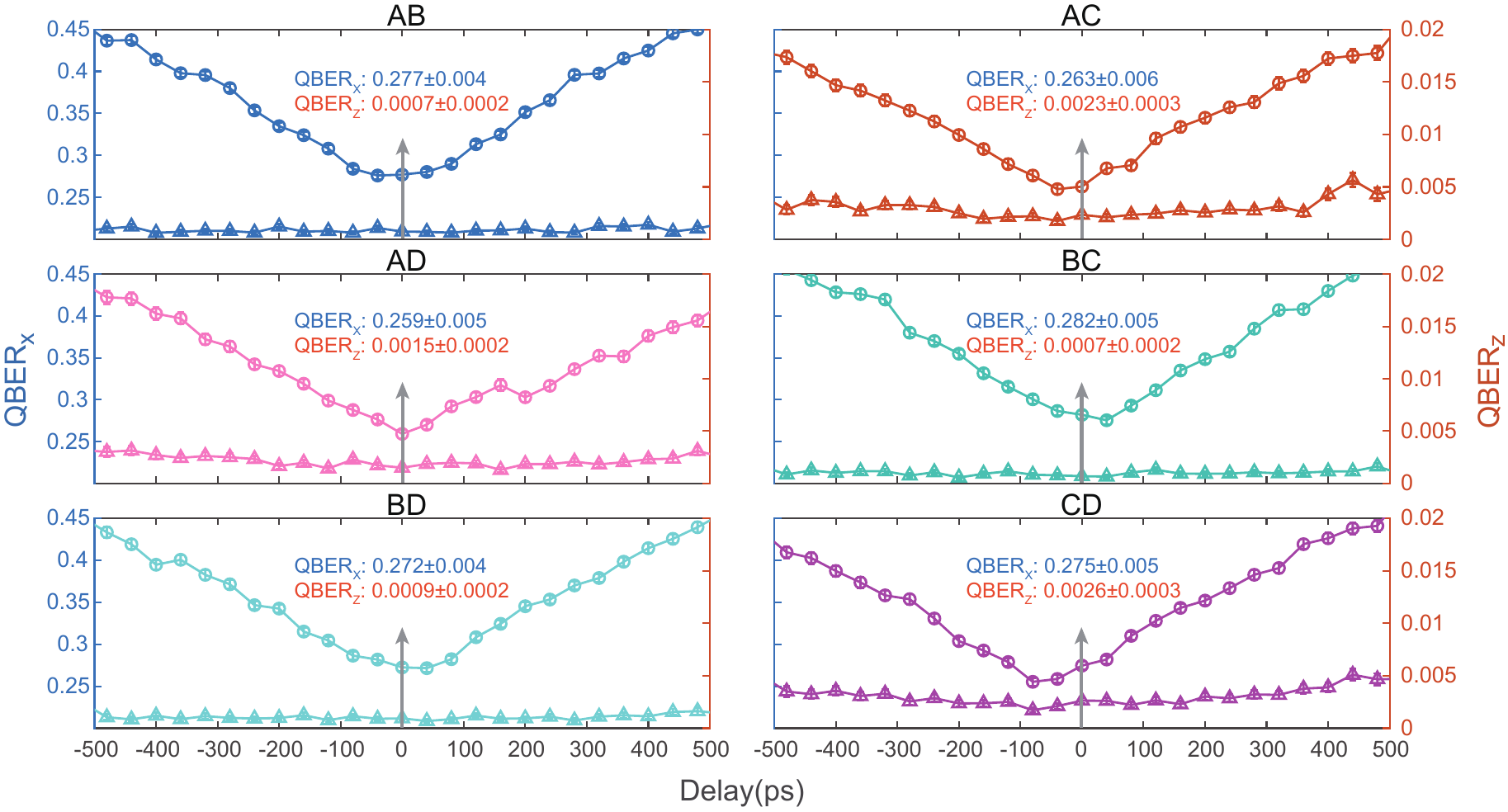}
    \caption{\label{Fig10} Quantum bit error rate of X basis and Z basis versus relative delay. $QBER_X$ data are marked by circles and $QBER_Z$ data are marked by triangles.}
\end{center}
\end{figure*}

We also measure the quantum bit error rate $QBER_X$ and $QBER_Z$ of different user pair as the function of relative delay. The mean photon number of qubits of Z basis and X basis are same, both are about 0.64. Every user produce 4 states ($\ket{0}$,$\ket{1}$,$\ket{+}$ and $\ket{-}$) randomly and send them to untrusted relay server for BSM. We scan the relative electrical delay of different AWG channels and record 20 seconds data at each delay point. Then we obtain the $QBER_X$ and $QBER_Z$ shown in Fig. \ref{Fig10}. The QBER in X basis at 0 delay point are $0.277 \pm 0.004$, $0.263 \pm 0.006$, $0.259 \pm 0.005$, $0.282 \pm 0.005$, $0.272 \pm 0.004$ and $0.275 \pm 0.005$ for user pair AB, AC, AD, BC, BD and CD, respectively. The QBER in Z basis at 0 delay are $0.0007 \pm 0.0002$, $0.0023 \pm 0.0003$, $0.0015 \pm 0.0002$, $0.0007 \pm 0.0002$, $0.0009 \pm 0.0002$ and $0.0026 \pm 0.0003$ for user pair AB, AC, AD, BC, BD and CD, respectively.

\subsection{Key rate analysis}\label{sec:three}
In practical quantum key distribution experiments, the number of pulses sent by the users is finite. In order to extract the security key, the statistical fluctuation effect should be taken into consideration. The yield and phase error rate of the single-photon components $s_{11}$ and $e^{\text{ph}}_{11}$ need to be estimated.    
We estimate the the expectation values from the experimental observations according to the Chernoff bound~\cite{curty2014finite,Yin2016,Zhang2017,Jiang2021}. Let $N_{lr}$ and $S_{lr}$ be the number of sent pulses and yield of pulses from source $l$, $r$$\in$\{$z$, $y$, $x$, $o$\}. Given the failure probability $\epsilon$, we can obtain the lower bound $\underline{S}_{l r}$ and upper bound $\bar{S}_{l r}$ of expectation values $\langle S_{lr}\rangle$: 
    \begin{equation}
        \underline{S}_{l r}=S_{l r} /\left(1+\delta_{l r}\right) \text { and } \bar{S}_{l r}=S_{l r} /\left(1-\delta_{l r}\right)
    \end{equation}
    in which 
    \begin{equation}
        \delta_{l r} = \delta\left(N_{l r} S_{l r}, \epsilon\right)=\frac{b+\sqrt{b^2+8 b N_{l r} S_{l r}}}{2 N_{l r} S_{l r}}
    \end{equation}
    with $b=-\ln(\epsilon/2)$. In addition, in order to get a tighter estimation, joint constraints among different variables are introduced. According to the method in Ref.~\cite{Zhou2016}, we can obtain the lower bound of $s_{11}$ in Z basis and upper bound of $e^{\text{ph}}_{11}$ as the function $\mathcal{H}$: 
    \begin{equation}
\mathcal{H}=a_{0}\langle S_{ox}\rangle+b_{0}\langle S_{xo}\rangle-a_{0}b_{0}\langle S_{oo}\rangle
    \end{equation}
    where $a_{0}=b_{0}=e^{-x}$.The secure key rate formula can be represented as the function of $\mathcal{H}$:
    \begin{equation}
        \mathcal{R}(\mathcal{H})=p_z p_z\left\{z^{2}e^{-2z} \underline{s}_{11}(\mathcal{H})\left[1-H\left(\bar{e}_{11}^{p h}(\mathcal{H})\right)\right]-f S_{z z} H\left(E_{z z}\right)\right\}
    \end{equation}
where $H(\cdot)$ is the Shannon binary entropy function, $f=1.16$ is the error correction efficiency, $S_{zz}$ and $E_{zz}$ are the observed yield and bit error rate for source $zz$, respectively.  Then the final secure key rate is the minimum of $\mathcal{R}(\mathcal{H})$ and the range of $\mathcal{H}$ can be calculated with same method in Ref.~\cite{Zhou2016}.

\subsection{Experimental results}\label{sec:four}

We run the MDI-QN four times continuously. In every run, only 3 user pairs' results are measured due to the limitation of available detectors. And a total of about $3\times 10^{12}$ pulse pairs are sent from each user side.

In Table \ref{tab:1}, we show the measurement data of total gains and error gains of Bell state $\ket{\Psi^-}$ at about 30 dB loss. The notion $N_{ij}$
denotes the number of pulse pairs sent out from $ij$ intensity combination. 
\begin{table}[!htbp]
\centering
\caption{List of detailed experimental results at the loss of about 30 dB.}
\label{tab:1}
\resizebox{\linewidth}{!}{
\begin{tabular}{| l | c | c | c || c | c | c || c | c | c || c | c | c |}
    \hline
User pair            & AB        & AC        & AD        & AB        & BC        & BD       & AC        & BC        & CD       & AD        & BD        & CD \\ \hline
Loss (dB)            & 30.6      & 30.9      & 30.3      & 30.7      & 30.9      & 30.5     & 30.9      & 30.0      & 29.3     & 30.8      & 30.4      & 30.1 \\ \hline
$N_{zz}S_{zz}$       & 87788209  & 89582267  & 92829111  & 84866989  & 81718178  & 97819037 & 80927353  & 99865391  & 118650729 & 90536935  & 99600084  & 96763342\\ \hline
$N_{zz}S_{zz}E_{zz}$ & 256301    & 128966    & 224336    & 310263     & 276488   & 197830   & 131921    & 301716    & 408261   & 350307     & 162378     & 165474 \\ \hline
$N_{yy}S_{yy}$       & 48055     & 50688     & 44783     & 44076     & 41118     & 48831    & 37498     & 46322     & 58217    & 51288     & 54175     & 53694 \\ \hline
$N_{xx}S_{xx}$       & 88556     & 90639     & 89536     & 97554     & 88367     & 103289   & 88997     & 97974     & 110892    & 84991     & 108011     & 95910 \\ \hline
$N_{xx}S_{xx}E_{xx}$ & 24926     & 25492     & 23222     & 26980     & 24187     & 26444    & 27602     & 28441     & 31205    & 23070    & 30985     & 27781 \\ \hline
$N_{yo}S_{yo}+N_{oy}S_{oy}$ & 13375 & 16899    & 14250   & 12694     & 10924     & 16540    & 12120     & 14847     & 21649     & 17511      & 17504     & 17622 \\ \hline
$N_{xo}S_{xo}+N_{ox}S_{ox}$ & 5923  & 4992  & 5888      & 6387      & 5210       & 5119     & 4756      & 6098      & 6739     & 5642      & 5651      & 6277 \\ \hline
$N_{oo}S_{oo}$              & 0     &  0    & 0         & 0         & 0          & 0         & 0        & 0         & 0        & 0         & 0         & 0\\ \hline
$s_{11}$          & $1.02\times10^{-4}$  & $1.54\times10^{-4}$   & $1.23\times10^{-4} $  & $1.38\times10^{-4} $ & $1.41\times10^{-4} $ & $2.15\times10^{-4}$ & $1.78\times10^{-4}$  & $1.53\times10^{-4}$   & $1.81\times10^{-4} $  & $1.02\times10^{-4} $ & $2.04\times10^{-4} $ & $1.24\times10^{-4}$ \\ \hline
$e^{ph}_{11}$     & 0.1455    & 0.1941     & 0.0734    & 0.1149    & 0.1565    & 0.1469 & 0.2291    & 0.1651     & 0.1463    & 0.1160    & 0.2005    & 0.1648\\ \hline
Key rate/ pulse   &$1.64\times10^{-6} $  & $2.32\times10^{-6} $  & $4.02\times10^{-6}  $ & $3.17\times10^{-6} $ & $2.34\times10^{-6}$  & $4.07\times10^{-6}$&$2.01\times10^{-6} $  & $2.33\times10^{-6} $  & $3.13\times10^{-6}  $ & $1.89\times10^{-6} $ & $2.95\times10^{-6}$  & $2.13\times10^{-6}$
\\
\hline
    
\end{tabular}}
\end{table}

In Table \ref{tab:2}, we show the comparison of state-of-the-art QKD network without trusted nodes. Our work has improved greatly in key rate. Moreover, it can support simultaneous communication, which is not available in other previous works.
\begin{table}[!htbp]
\centering
\caption{Comparison of state-of-the-art QKD network without trusted nodes.}
\label{tab:2}
\begin{threeparttable}
\resizebox{\linewidth}{!}{
\begin{tabular}{ l  c  c  c  c  }
\hline
\hline
Reference & Protocol & Number of users & Key rate (bps) & Simultaneous communication\\
\hline
Wengerowsky \textit{et al.}\cite{wengerowsky2018} & BBM92 & 4 & 3 $\sim$ 15 (estimated by authors) & No \\ 
\hline
Joshi \textit{et al.}\cite{joshi2020} & BBM92 & 8 & 58 $\sim$ 304 (20 m fiber) & No\\ 
\hline
Tang \textit{et al.}\cite{PhysRevX.6.011024} & MDI & 3 & 16.5 $\sim$ 38.8 ($\sim$ 20 dB)& No\\ 
\hline
This work  & MDI & 4 & 164$\sim$407 ($\sim$ 30 dB)& Yes\\ 
\hline 
\hline
\end{tabular}
}
\end{threeparttable}
\end{table}

\subsection{Hardware efficient protocol for a fully-connected MDI-QKD network}\label{sec:five}

In this section, we propose a hardware efficient protocol, using the time-division multiplexing (TDM), to reduce our network’s detector resource consumption from $O(n^2$) to $O(n$) scaling for an n-user network. The number of wavelengths and the corresponding encoders is proportional to $\lceil log_{2}n\rceil$. The number of BSM modules required is $2^{\lceil log_{2}n\rceil}-1$.

\begin{figure*}[htbp]
\begin{center}
    \includegraphics[width=1\textwidth]{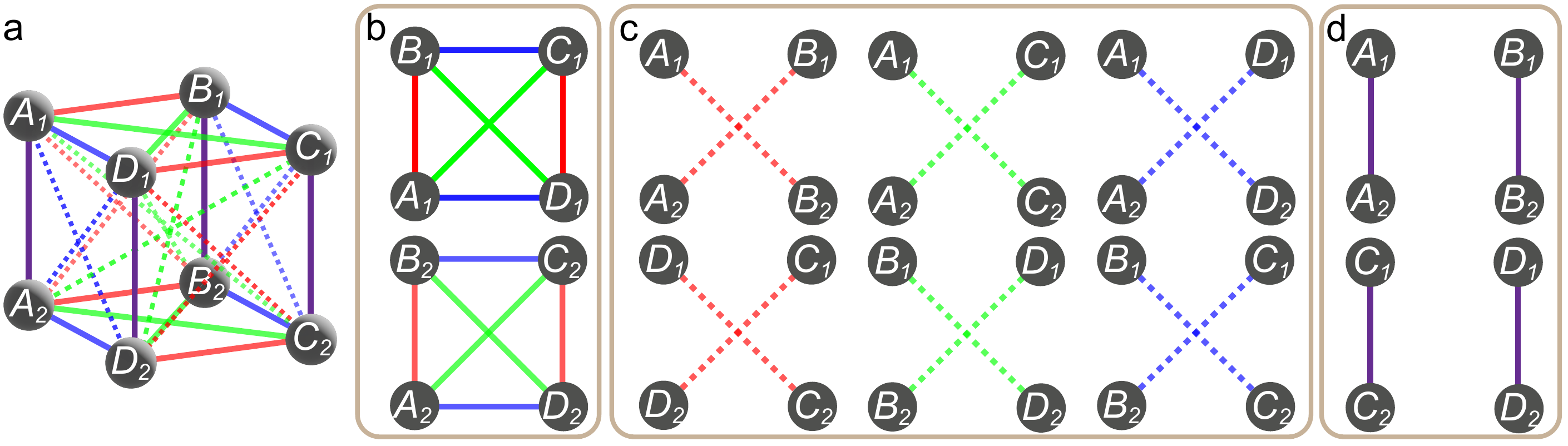}
    \caption{\label{Fig11} \textbf{a}, Topology of an eight-user fully connected MDI-QKD network. \textbf{b}, Internal communication between subnets. \textbf{c}, Cross communication between two subnets. \textbf{d}, Additional communication.}
\end{center}
\end{figure*}
In the following, we will illustrate the scale of our work using an eight-user communication network as an example. The topology of a eight-user fully connected network is shown in Fig. \ref{Fig11}\textbf{a}, which consists of two identical four-user subnets ($A_1-B_1-C_1-D_1$ and $A_2-B_2-C_2-D_2$). The eight-user system only needs to add one additional light wavelength and one additional BSM module, comparing to the resource consumption for our original four-user network.  
Each complete communication process of the eight-user network can be temporally divided into four time bins, $t_1-t_2-t_3-t_4$. The internal communication of the two subnets (Fig. \ref{Fig11}\textbf{b}) is implemented at the $t_1$ and $t_4$ time bins with six BSM modules (\textbf{BSM 1$\sim$6} in Table \ref{tab:3}), respectively. The cross-communication between the two subnets (Fig. \ref{Fig11}\textbf{c}) is implemented at $t_2$ and $t_3$ time, respectively. The additional communication (Fig. \ref{Fig11}\textbf{d}) between the four user pairs is performed on the a new BSM module (\textbf{BSM 7} in Table S3) with a new light wavelength.

\begin{table}[!htbp]
    \centering
    \centering
    \caption{The specific communication process of eight-user MDI-QKD network}
    \label{tab:3}
    \renewcommand\arraystretch{1}
    \begin{tabular}{|c|c|c|c|c|c|c|c|c|}
        \hline
        & \textbf{Bin}   & {\color{red} \textbf{BSM 1}}     & {\color{green} \textbf{BSM 2}}     & {\color{blue} \textbf{BSM 3}}     & {\color{blue} \textbf{BSM 4}}     & {\color{green} \textbf{BSM 5}}     & {\color{red} \textbf{BSM 6}}     & {\color{violet} \textbf{BSM 7}}     \\ \hline
        \textbf{Subnet 1}                          & $t_1$ & {\color{red} $A_1-B_1$} & {\color{green} $A_1-C_1$} & {\color{blue} $A_1-D_1$} & {\color{blue} $B_1-C_1$} & {\color{green} $B_1-D_1$} & {\color{red} $C_1-D_1$} & {\color{violet} $A_1-A_2$} \\ \hline
        \multirow{2}{*}{\textbf{Cross subnet 1-2}} & $t_2$ & {\color{red} $A_1-B_2$} & {\color{green} $A_1-C_2$} & {\color{blue} $A_1-D_2$} & {\color{blue} $B_1-C_2$} & {\color{green} $B_1-D_2$} & {\color{red} $C_1-D_2$} & {\color{violet} $B_1-B_2$} \\ \cline{2-9} 
        & $t_3$ & {\color{red} $A_2-B_1$} & {\color{green} $A_2-C_1$} & {\color{blue} $A_2-D_1$} & {\color{blue} $B_2-C_1$} & {\color{green} $B_2-D_1$} & {\color{red} $C_2-D_1$} & {\color{violet} $C_1-C_2$} \\ \hline
        \textbf{Subnet 2}                          & $t_4$ & {\color{red} $A_2-B_2$} & {\color{green} $A_2-C_2$} & {\color{blue} $A_2-D_2$} & {\color{blue} $B_2-C_2$} & {\color{green} $B_2-D_2$} & {\color{red} $C_2-D_2$} & {\color{violet} $D_1-D_2$} \\ \hline
    \end{tabular}
\end{table}

We can further connect two eight-user networks, and extend to a 16-user fully connected MDI-QKD network consisting of four subnets. The 16-user network needs six light wavelengths and fifteen BSM modules. Each complete communication process can be divided into eight time bins. The time multiplexing approach described above can be further extended to more users. For an n-user fully connected MDI-QKD network, the time-division multiplexing method reduces the consumption of encoding and detection resources. The number of required encoders of transmitter for different wavelengths per user is: $3\lceil\log_2{n} \rceil/2$, if $\lceil\log_2{n}\rceil$ is even; $3\lceil\log_2{n} \rceil/2-1/2$, if $\lceil\log_2{n}\rceil$ is odd. The number of BSM modules network needed is $2^{\lceil\log_2{n}\rceil}-1$. So, the number of required encoders of transmitter per user is $O(\log_2{n})$ and the number of BSM modules is $O(n)$. However, in this scheme, to add more users may reduce the operation capacity of the network.

\begin{figure*}[htbp]
\begin{center}
    \includegraphics[width=1\textwidth]{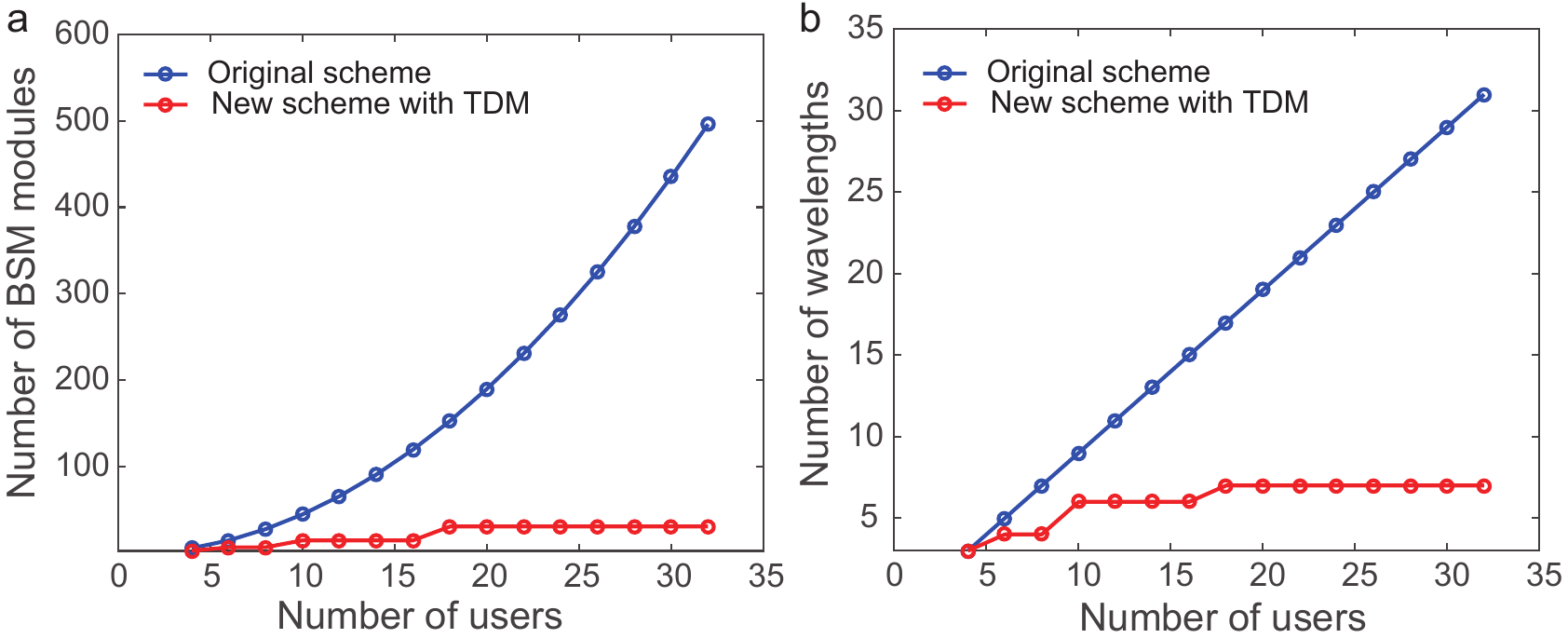}
    \caption{\label{Fig12} \textbf{a}, The BSM modules required for original scheme and new scheme with TDM . \textbf{b}, The wavelengths required for original scheme and new scheme with TDM.}
\end{center}
\end{figure*}

In Fig. \ref{Fig12}, we compared this new scheme with our original MDI-QN scheme in main text. In the new scheme with TDM, the consumption of required encoder resource of transmitter per user decreases from $O(n)$ to $O(\log_2{n})$, and the consumption of required detector resource of the network decreases from $O(n^2)$ to $O(n)$. Note that TDM approach reduces the duty cycle of the system and the communications between different pairs of users are allowed for a complete cycle (T=t1’ -t1=tm’ -tm, the period of each time bin). One can improve the key rate in other ways, such as increasing the system clock rate~\cite{Wei2020,Woodward2021} and using high-speed SNSPD to improve BSM efficiency~\cite{Zheng2021}.

\subsection{The linewidth measurement of DKS comb source}\label{sec:six}

\begin{figure*}[htbp]
\begin{center}
    \includegraphics[width=0.8\textwidth]{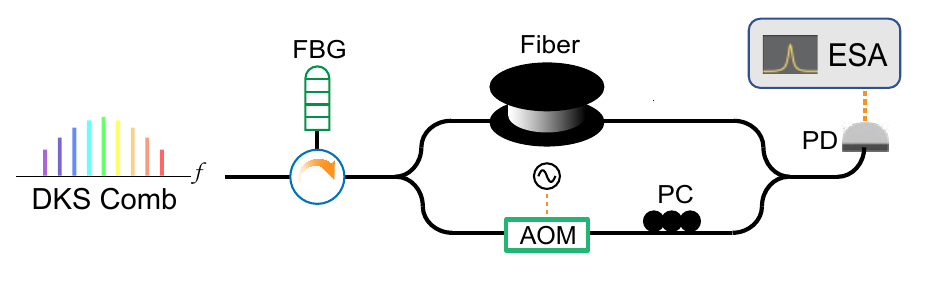}
    \caption{\label{Fig13} Schematic diagram of the delayed self-heterodyne method. FBG : 10 GHz fiber bragg grating;  AOM: acousto-optic modulator; PC : polarization controller; Fiber : long optical fiber of 5 km;  PD : photodiode; ESA : electrical spectrum analyzer.}
\end{center}
\end{figure*}

The delayed self-heterodyne is a refinement approach that incorporates an acousto-optic modulator (AOM) aimed at enhancing the stability of the test system. The structural model is illustrated in Fig. \ref{Fig13}. The entire test structure is based on the fibre Mach-Zehnder interferometer. The light source is then separated into two beams by means of a fibre coupler, and these two beams of light are passed through the upper and lower arms of the Mach-Zehnder interferometer. The light is transmitted in one arm through a disc of optical fibre to achieve the required time delay. The other arm is equipped with an acousto-optic modulator to move the comb line frequency through the operation. The light on the two arms of the interferometer is passed to the photodetector through the coupler, and then into the electrical spectrum analyzer to realise data acquisition and visualization.

This method involves the addition of an acousto-optic modulator to the delay self-heterodyne setup, with the purpose of reducing the influence of various low-frequency noises and direct current signals. The 10 GHz bandwidth fiber Bragg grating (FBG) filter is used to clean the frequency spectra of the DKS comb, thus obtaining a single comb line. The acousto-optic modulator operates at a modulation frequency of 100 MHz. The utilization of 5 km-long optical fiber is important to satisfy the delay condition, thereby facilitating the reading of the -10 dB bandwidth signal of the spectrum analyzer. The implementation of Lorentz linear fitting enables a linewidth of approximately 0.661 MHz. The line width of a single comb line, which is equivalent to $1/2\sqrt{9}$ of the fitting, is about 0.1101 MHz . In Fig. \ref{Fig14} shown below, we obtain 1.4 MHz and 0.11 MHz for the chaotic state and soliton state comb lines, indicating excellent coherent property of soliton combs.

\begin{figure*}[htbp]
\begin{center}
    \includegraphics[width=0.8\textwidth]{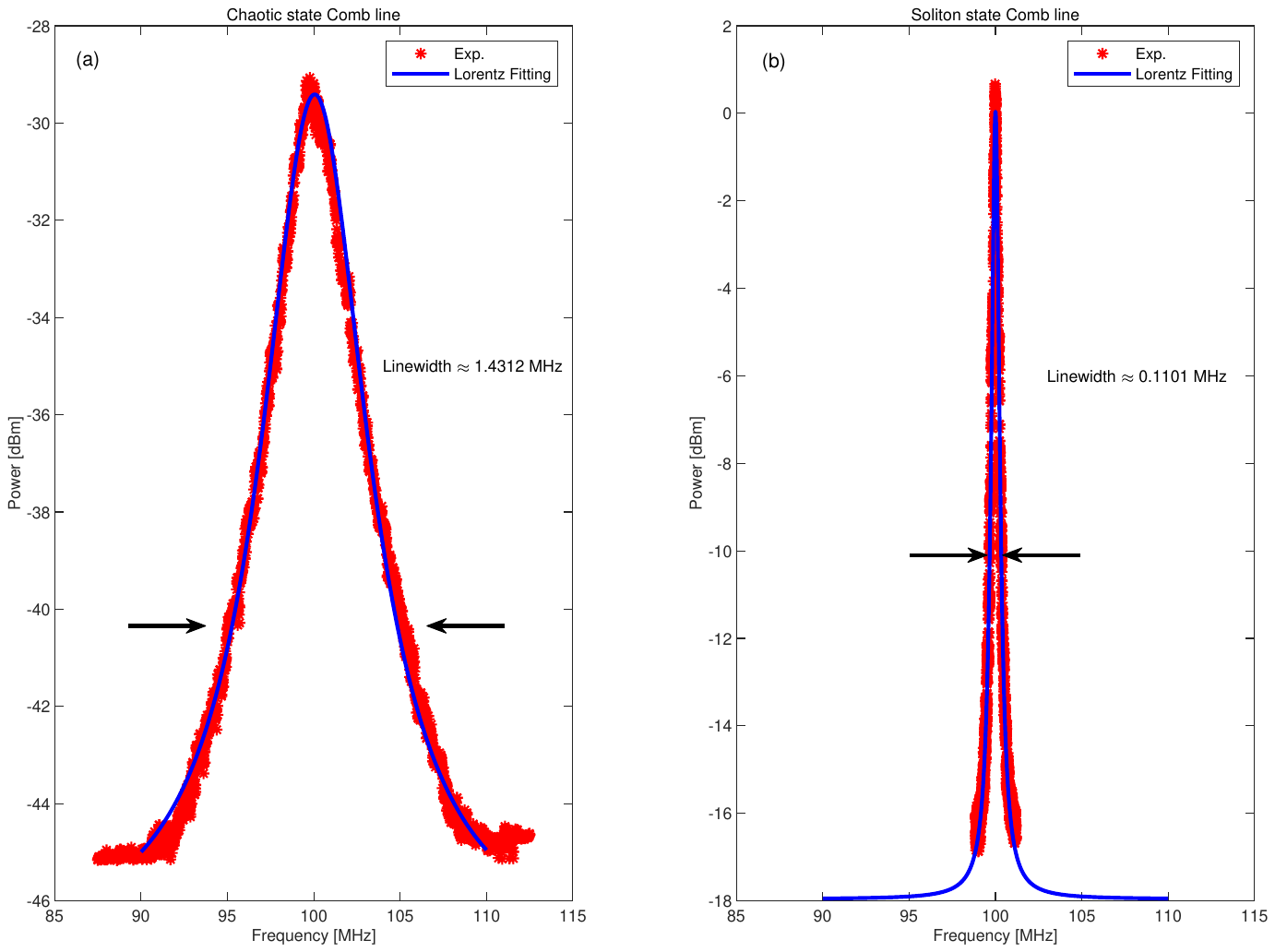}
    \caption{\label{Fig14} Plot and Lorentz fitting of the experimental linewidth data of the different comb line from the electrical spectrum analyser. \textbf{a}, The measured data of the chaotic state. \textbf{b}, The measured data of the soliton state.}
\end{center}
\end{figure*}

\clearpage
\end{document}